\documentclass[prd,superscriptaddress,amsfonts,amssymb,amsmath,showpacs,twocolumn]{revtex4-2}
\usepackage{bm}
\usepackage{amsfonts}
\usepackage{latexsym}
\usepackage{graphicx}
\usepackage{amsmath}
\usepackage{palatino}
\usepackage{mathpazo}
\usepackage{textcomp}
\usepackage{float}
\usepackage{booktabs}
\usepackage{dcolumn}
\usepackage{booktabs}
\usepackage{multirow}
\usepackage{hyperref}
\hypersetup{colorlinks,citecolor=blue}
\usepackage{amsmath}
\usepackage{xcolor}
\usepackage{orcidlink}
\usepackage[caption=false]{subfig}
\usepackage{commath}
\def\jnl@style{\it}
\def\aaref@jnl#1{{\jnl@style#1}}

\def\aaref@jnl#1{{\jnl@style#1}}

\def\aj{\aaref@jnl{AJ}}                   
\def\apj{\aaref@jnl{ApJ}}                 
\def\apjl{\aaref@jnl{ApJ}}                
\def\apjs{\aaref@jnl{ApJS}}               
\def\apss{\aaref@jnl{Ap\&SS}}             
\def\aap{\aaref@jnl{A\&A}}                
\def\aapr{\aaref@jnl{A\&A~Rev.}}          
\def\aaps{\aaref@jnl{A\&AS}}              
\def\mnras{\aaref@jnl{Mon.~Not.~Roy.~Astron.~Soc.}}             
\def\prd{\aaref@jnl{Phys.~Rev.~D}}        
\def\prc{\aaref@jnl{Phys.~Rev.~C}}  
\def\prl{\aaref@jnl{Phys.~Rev.~Lett.}}    
\def\qjras{\aaref@jnl{QJRAS}}             
\def\skytel{\aaref@jnl{S\&T}}             
\def\ssr{\aaref@jnl{Space~Sci.~Rev.}}     
\def\zap{\aaref@jnl{ZAp}}                 
\def\nat{\aaref@jnl{Nature}}              
\def\aplett{\aaref@jnl{Astrophys.~Lett.}} 
\def\apspr{\aaref@jnl{Astrophys.~Space~Phys.~Res.}} 
\def\physrep{\aaref@jnl{Phys.~Rep.}}      
\def\physscr{\aaref@jnl{Phys.~Scr}}       
\def\commat{\aaref@jnl{Comm.~Math.~Phys.}}              
\def\science{\aaref@jnl{Science}}               
\def\cqg{\aaref@jnl{Classical Quant.~Grav.}}            
\def\jpcs{\aaref@jnl{JPCS}}                                     
\def\ijmpd{\aaref@jnl{Int.~J.~Mod.~Phys.~D}}                    
\def\grg{\aaref@jnl{Gen.~Relat.~Gravit.}}               
\def\rpp{\aaref@jnl{Rep.~Prog.~Phys.}}          
\def\npa{\aaref@jnl{Nucl.~Phys.~A}}        
\def\lrr{\aaref@jnl{Living Rev.~Rel.}}                   
\def\jcap{\aaref@jnl{J.~Cosmology Astropart.~Phys.}}    
\def\rmp{\aaref@jnl{Rev.~Mod.~Phys.}}   
\def\epjc{\aaref@jnl{Eur.~Phys.~J.~C}}

\allowdisplaybreaks[1]
\renewcommand{\arraystretch}{1.1}
\begin{document}

\color{black}       

\title{Scalar field dark energy: Insights into cosmological evolution and black hole accretion}

\author{M. Koussour\orcidlink{0000-0002-4188-0572}}
\email[Email: ]{pr.mouhssine@gmail.com}
\affiliation{Quantum Physics and Magnetism Team, LPMC, Faculty of Science Ben
M'sik,\\
Casablanca Hassan II University,
Morocco.} 

\author{Y. Sekhmani\orcidlink{0000-0001-7448-4579}}
\email[Email: ]{sekhmaniyassine@gmail.com}
\affiliation{Ratbay Myrzakulov Eurasian International Centre for Theoretical Physics, Astana 010009, Kazakhstan.}

\author{Alnadhief H. A. Alfedeel\orcidlink{0000-0002-8036-268X}}%
\email[Email: ]{aaalnadhief@imamu.edu.sa}
\affiliation{Department of Mathematics and Statistics, Imam Mohammad Ibn Saud Islamic University (IMSIU),\\
Riyadh 13318, Saudi Arabia.}
\affiliation{Department of Physics, Faculty of Science, University of Khartoum, P.O. Box 321, Khartoum 11115, Sudan.}
\affiliation{Centre for Space Research, North-West University, Potchefstroom 2520, South Africa.}

\author{F. Awad}%
\email[Email: ]{awad.fga@gmail.com}
\affiliation{Department of Mathematics, Omdurman Islamic Univerity, Khartoum, Sudan.}
\affiliation{Alneelain Center for Mathematical Sceinces, Alaneelain University, Kartoum, Sudan.}

\author{N. Myrzakulov\orcidlink{0000-0001-8691-9939}}
\email[Email: ]{nmyrzakulov@gmail.com}
\affiliation{L. N. Gumilyov Eurasian National University, Astana 010008,
Kazakhstan.}
\affiliation{Ratbay Myrzakulov Eurasian International Centre for Theoretical
Physics, Astana 010009, Kazakhstan.}


\date{\today}

\begin{abstract}
We propose a novel approach to parameterize the equation of state for Scalar Field Dark Energy (SFDE) and use it to derive analytical solutions for various cosmological parameters. Using statistical MCMC with Bayesian techniques, we obtain constraint values for the model parameters and analyze three observational datasets. We find a quintessence-like behavior for Dark Energy (DE) with positive values for both model parameters $\alpha$ and $\beta$. Our analysis of the $CC$+$BAO$+$SNe$ datasets reveals that the transition redshift and the current value of the deceleration parameter are $z_{tr}=0.73_{-0.01}^{+0.03}$ and $q_{0}=-0.44_{-0.02}^{+0.03}$, respectively. We also investigate the fluid flow of accretion SFDE around a Black Hole (BH) and analyze the nature of the BH's dynamical mass during accretion, taking into account Hawking radiation and BH evaporation. Our proposed model offers insight into the nature of DE in the Universe and the behavior of BHs during accretion.
\end{abstract}

\maketitle

\section{Introduction}
\label{sec1}

We are presently situated in a unique epoch of cosmic history, where the expansion of the Universe is not decelerating but accelerating. The exact cause of this acceleration, however, remains unknown. In the  background of General Relativity Theory (GR), the late-time acceleration of the Universe can be explained by the inclusion of Dark Energy (DE) density, in addition to matter density, in Einstein's Field Equations (EFEs). There have been several studies on this topic \cite{DDE}. On the other hand, some modified theories of gravity (MTG) have also been proposed that can explain the present acceleration of the Universe without the inclusion of DE. Nonetheless, the nature of DE and its role in the acceleration of the Universe remains one of the most significant unanswered questions in modern physics. Observationally, the late-time acceleration of the Universe has been extensively studied using various techniques, including the luminosity distance of Type Ia Supernovae (SNe) \cite{Riess, Perlmutter}. Apart from SNe observations, other independent observations such as Baryon Acoustic Oscillations (BAO) \cite{D.J., W.J.}, the Cosmic Microwave Background (CMB) \cite{R.R., Z.Y.}, the Large Scale Structure (LSS) \cite{T.Koivisto, S.F.} and the recent Planck collaboration \cite{Planck2020} have also provided evidence for the present acceleration of the Universe. According to observational estimations, dark matter, and hypothetical DE make up the majority of the Universe. The essential characteristics of these enigmatic components are still mostly unknown despite substantial research into their physical nature. The simplest method to describe the observable acceleration of the Universe is to modify EFEs to include the small cosmological constant $\Lambda$. The energy density of $\Lambda$ is equal to its pressure $p_{\Lambda}=-\rho_{\Lambda}$, which is negative \cite{Sahni0}. This approach is known as the $\Lambda$CDM model and has been remarkably successful in explaining a wide range of cosmological observations. However, it is not without its challenges, and many researchers continue to explore alternative explications for the acceleration of the Universe. In addition to the theoretical problems of fine-tuning and cosmic coincidence \cite{dalal/2001, weinberg/1989}, the $\Lambda$CDM model also faces the $H_{0}$ tension, which is currently one of the major challenges within this paradigm. The $H_{0}$ tension arises due to the discrepancy between the measurements of the Hubble constant obtained from the early Universe observations, such as the CMB, and the late Universe observations, such as the distance measurements of nearby SNe and the Hubble Space Telescope \cite{HT0,HT1,HT2,HT3,Planck2020}. The tension has led to various proposals for new physics beyond the standard $\Lambda$CDM model, such as modifications to the DE equation of state (EoS), the inclusion of extra relativistic species, and even modifications to the theory of gravity itself \cite{Vagnozzi}. 

One alternative approach to describing the late-time acceleration of the Universe is to assume the Einstein-Hilbert Lagrangian as a general function of various scalar quantities (or the so-called MGT), such as the Ricci scalar $R$, the torsion scalar $T$, or the non-metricity scalar $Q$. In these MGT, it is possible to explain the Universe's expansion without the need for DE and instead relies on the modification of the gravitational interaction at large scales. Such theories have been proposed as alternatives to the $\Lambda$CDM model and have been studied extensively in recent years \cite{fR1, fR2,fT1,fT2,fQ1, fQ2,fQ3, fQ4, fQ5, fQ6, fQ7}. However, there are challenges in constructing viable MGT models that can successfully explain the observational data, and the current status of these theories is still a subject of ongoing research and debate.

Moreover, the scalar fields $\phi$ with a time or redshift varying EoS are a popular approach to explaining the late-time acceleration of the Universe. These scalar fields emerge negative pressure as they slowly roll down their scalar potential $V(\phi)$ \cite{SF0,SF1}. The quintessence scalar field introduced the idea of tracker potentials, which can avoid fine-tuning and coincidence problems \cite{SF2,SF3}. Ref. \cite{SF4} discusses the concept of tracker fields in quintessence theory as a means of explaining the current acceleration of the Universe. The article presents the idea of integrated tracking, which suggests that tracker potentials follow a definite path of the Universe's evolution in accordance with observational constraints. The concept of tracker fields is useful in avoiding the fine-tuning and coincidence problems encountered by other models. Various applications of the time-varying EoS parameter have been discussed in literature \cite{Ratra,Peebles,Wetterich,Carroll,Y.Fujii,Phantom1,Phantom2,T.Chiba,C.Arm.,Kamenshchik,M.C.,A.Y.,tachyon}. Recently, Singh et al. explored the properties and behavior of DE in the FLRW cosmology using the EDSFD parametrization \cite{Singh}. Pacif et al. employed a scalar field source to examine the Universe's late-time acceleration and proposed a scalar field model that is consistent with observational data \cite{Pacif}. Bairagi investigated the characteristics of DE models in the setting of non-canonical scalar fields within the Einstein-Aether Gravity framework \cite{Bairagi}. Debnath and Bamba studied a non-canonical scalar field in a D-dimensional fractal universe to examine the behavior of DE models \cite{Debnath}. Kar et al. examined the relationship between the $f(Q)$ gravity model and the Dirac-Born-Infeld Scalar Field DE (SFDE) model \cite{Kar}. In the setting of $f(R,T)$ gravity, Sharma et al. explored how the scalar field models of the Barrow holographic DE behaved. \cite{Sharma}. The paper by Koussour \cite{MK} focuses on the accretion process of a Black Hole (BH) in the context of an SFDE model. The author considers a non-rotating BH and analyzes the energy and momentum transport in the surrounding SFDE medium.

In the present paper, we study the behavior of a DE model with a scalar field in a spatially flat FLRW (Friedmann-Lemaître-Robertson-Walker) Universe. To achieve this, we consider a novel approach to parameterize the EoS for SFDE and use it to obtain analytical solutions for various cosmological parameters. The parametrization method of DE plays a crucial role in studying the large-scale properties and evolution of the Universe. They provide a way to model the behavior of DE in different cosmological contexts, and they can be used to test the compatibility of different observational datasets with various DE models. There are many different parametrization methods available in the literature, each with its own strengths and weaknesses. Some popular methods include EoS parametrization, redshift-based parametrization, and principal component analysis. For more information on these and other parametrization methods, see the Refs. \cite{Cunha, Mortsell, Pacif1}. Here, we use the statistical MCMC method with the Bayesian technique to determine the constraint values for the parameters. We analyze the results using three observational datasets: the cosmic chronometers, BAO, and SNe datasets.

The most significant of the numerous predictions of GR is the prediction of BH. The tendency of BHs to accrete is an obvious outcome, and various aspects of accretion in BHs are investigated \cite{1}. A number of astrophysicists describe accretion as the inflow of matter into the center of an object where the gravitational forces are extremely strong or are moving into the center of the mass. The accretion process on compact objects was first examined by Bondi  in a Newtonian context \cite{2}. Michel \cite{3} found a similar problem regarding the relativistic results from which the work will take care of the analysis of the acrretion onto Schwarzschlid BH in the context of GR. The attempts to better analyze the process of accretion for charged BHs are studied in \cite{4,5} . Another interesting process is what is called BH evaporation. The BH evaporation theory is a kind of landmark in theoretical physics. Indeed, investigations of the evaporation process  have been a rewarding area of research in theoretical physics ever since the proves of Hawking radiation. Within the framework of quantum field theory on a curved space-time background, BHs emit thermal radiation, affected by the scattering effects from the space-time geometry. The backreaction of the radiation flux leads to the evaporation of the BH, which should lose all its mass and angular momentum.

The paper is structured as follows: In Sec. \ref{sec2}, we introduce the field equations of GR and scalar field coupling. Next, in Sec. \ref{sec3}, we present the cosmological solutions obtained using a novel approach to parameterize the EoS for SFDE and derive the corresponding cosmological parameters. Then, in Sec. \ref{sec4}, we use the $CC$+$BAO$+$SNe$ datasets to determine the best-fit values of the model parameters. Moreover, in Sec. \ref{sec5} we examine the behavior of the cosmological parameters for the model parameters that are constrained by the observational datasets. In Sec. \ref{sec6}, we apply statefinder and $Om(z)$ diagnostics to distinguish our SFDE cosmological model from other DE models. Sec. \ref{sec7} of our study investigates the dynamical mass of the BH as it accretes the fluid flow, taking into account the Hawking radiation and the subsequent evaporation of the BH. Lastly, we discuss our findings and conclude in Sec. \ref{sec8}.

In this paper, the convention of setting $8\pi G=c=1$ has been consistently used.

\section{Understanding DE: Exploring the Field Equations with Scalar Field}

\label{sec2}

The action governing a Universe with spacetime curvature $R$, comprising a scalar field $\phi$ with a potential energy function $V(\phi)$ and ordinary matter, can be expressed as the sum of the so-called Einstein-Hilbert action, the scalar field action, and the matter action,%
\begin{equation}
S = \int\sqrt{-g} d^{4}x{\left[\frac{1}{2\kappa }R + \frac{1}{2}\partial _{\mu }\phi \partial ^{\mu }\phi -V\left( \phi
\right) + L_{m}\right]}.
\label{Action}
\end{equation}%

Here, the symbol $L_{m}$ refers to the Lagrangian density of the matter, which is assumed to consist of a pressureless perfect fluid. This Lagrangian accounts for the behavior of ordinary matter in the Universe and how it interacts with the scalar field and curvature of spacetime. The determinant of the metric, which describes the relationship between spacetime coordinates, is denoted by the symbol $'g'$. In this context, the constant $\kappa$ is defined as $\kappa = \frac{8\pi G}{c^4}$, where $G$ is the gravitational constant and $c$ is the speed of light. In the above equation, the Einstein-Hilbert action represents the curvature of spacetime due to matter and energy, whereas the scalar field action  represents the behavior of the scalar field $\phi$. The potential $V(\phi)$ signifies the energy density of the scalar field, which can vary depending on its value at a given point in spacetime. The action provides a way to calculate the equations of motion for the scalar field and matter, and their combined effect on the curvature of spacetime. These equations sometimes referred to as the field equations with the scalar field, can assist us in comprehending DE behavior, which is assumed to be related to the properties of the scalar field.

In this study, we investigate a spatially flat FLRW spacetime, which represents a homogeneous and isotropic Universe, the metric can be expressed  in the following form:

\begin{equation}
ds^{2}=-dt^{2}+a^{2}(t)\left[ dr^{2}+r^{2}\left( d\theta ^{2}+\sin
^{2}\theta d\psi ^{2}\right) \right] .
\label{FLRW}
\end{equation}
where $t$ represents the cosmic time, $a(t)$ is the scale factor of the Universe, which describes how its size increases over time, $r$, $\theta$, and $\psi$ are the co-moving radial and angular coordinates, respectively. The Ricci scalar associated with the line element given above can be calculated as $R=-6(\dot{H}+2H^{2})$, where $\dot{H}$ represents the time derivative of the Hubble parameter $H$, which is defined as the rate of expansion of the Universe. The Hubble parameter can be expressed in terms of the scale factor as $H = \frac{\dot{a}}{a}$.

In addition, we investigate a perfect fluid in the Universe, which is assumed to have a certain energy density, pressure, and velocity. The energy-momentum tensor for the fluid can be written as,
\begin{equation}
T_{\mu \nu }=(\rho +p)u_{\mu
}u_{\nu }+pg_{\mu \nu },
\label{EMT}
\end{equation}
where $\rho$ is the energy density of the cosmic fluid, $p$ is its pressure, and $u_{\mu}$ is the four-velocity vector of the fluid. In the context of the field equations with a scalar field, the energy-momentum tensor for the perfect fluid is one of the sources that contribute to the curvature of spacetime. Here, we consider the Universe as consisting of two distinct components: pressureless matter and a scalar field. The pressureless matter, also known as dust, refers to matter that does not exert any pressure and is only subject to gravitational interactions. The scalar field, on the other hand, is a type of field that permeates all of space and has an energy density and pressure that vary with time.

By inserting the metric in Eq. (\ref{FLRW}) into the field equations with a scalar field, we can derive a set of equations that govern the dynamics of the scalar field and matter for a flat Universe as,%
\begin{equation}
3H^{2}=\rho_{m}+\rho_{\phi},  \label{F1}
\end{equation}%
and%
\begin{equation}
2{\dot{H}}+3H^{2}=-p_{\phi}.  \label{F2}
\end{equation}

Here, the total (or effective) energy density and pressure of the Universe can be written as the sum of the energy density and pressure of the matter, $\rho_{m}$ and $p_{m}=0$, and the energy density and pressure of the scalar field, $\rho_{\phi}$ and $p_{\phi}$, respectively. We denote the total energy density and pressure as $\rho_{T}=\rho_{m}+\rho_{\phi}$ and $p_{T}=p_{\phi}$, respectively. Moreover, the energy density $\rho _{\phi }$ and pressure $p_{\phi }$ of the scalar field can be expressed as functions of the scalar field's potential $V(\phi)$, and its kinetic energy $\frac{1}{2}\overset{.}{\phi }^{2}$ \cite{Barrow1,Barrow2},
\begin{equation}
\rho _{\phi }=\frac{1}{2}\overset{.}{\phi }^{2}+V\left( \phi \right) ,
\label{rho_phi}
\end{equation}%
and%
\begin{equation}
p_{\phi }=\frac{1}{2}\overset{.}{\phi }^{2}-V\left( \phi \right) .
\label{p_phi}
\end{equation}

The dynamics of the scalar field and matter field are governed by conservation equations, which describe the conservation of energy and momentum over time,
\begin{equation}
\overset{.}{\rho }_{m}+3\rho _{m}H=0,  \label{cm}
\end{equation}%
\begin{equation}
\overset{.}{\rho }_{\phi }+3\left( \rho _{\phi }+p_{\phi }\right) H=0.
\label{cf}
\end{equation}

By solving the equation mentioned above (\ref{cm}), we can obtain an expression for the energy density of normal matter $\rho _{m}$,%
\begin{equation}
\rho _{m}=\rho _{m0}\left( 1+z\right) ^{3},
\label{rhom}
\end{equation}%
where $\rho_{m0}$ in the above equation is an integration constant, which represents the energy density of normal matter at the present epoch. This parameter allows us to calibrate our models to match the observed properties of the Universe. The redshift parameter, denoted by $z$, is defined as $z=\frac{a_{0}}{a(t)}-1$ (where $a_{0}$ represents the current value of the scale factor, which is conventionally set to 1) and provides a way to relate the observed properties of celestial objects to their distance from us. By incorporating this parameter into our analysis, we can better understand the evolution of the Universe over time and its relationship to the properties of matter and energy.

It is also useful to write the derivative of the $H$ with respect to cosmic time in terms of redshift,
\begin{equation}
\overset{.}{H}=-\left(
1+z\right) H\left( z\right) \frac{dH\left( z\right) }{dz}.
\label{t_z}
\end{equation}

The behavior of the SFDE can be described by its equation of state (EoS) parameter $\omega _{\phi}$, which relates to the field's pressure and energy density i.e. $\omega _{\phi }=\frac{p_{\phi }}{\rho _{\phi }}$. From Eqs. (\ref{F1})-(\ref{p_phi}), the EoS parameter for the SFDE can be expressed as follows:
\begin{equation}
\omega _{\phi}=-\frac{\left( 2\overset{.}{H}+3H^{2}\right) }{3H^{2}-\rho
_{m}}=\frac{\overset{.}{\phi }^{2}-2V\left( \phi \right) }{\overset{.}{\phi }%
^{2}+2V\left( \phi \right) }.
\label{EoS}
\end{equation}

By combining the EoS parameter contributions of the scalar field and normal matter, we can obtain the total equation of state (EoS) parameter i.e. $\omega _{T}=\frac{p_{T}}{\rho _{T}}$ as,
\begin{equation}
\omega _{T}=\frac{p_{\phi }}{\rho _{m}+\rho
_{\phi }}=-\frac{\left( 2\overset{.}{H}+3H^{2}\right) }{3H^{2}}.
\end{equation}%

This parameter describes the overall relationship between the total pressure and energy density of the Universe, and can provide valuable insights into its behavior and evolution. From Eqs. (\ref{F1}) and (\ref{F2}),  we have only a system of two independent equations and three unknown parameters, namely, $H$, $\phi$ and $V(\phi)$. Therefore, an additional input is required to solve the system of equations.

\section{Novel approach to parameterizing the EoS parameter}
\label{sec3}

It is important to explore models beyond the cosmological constant to better understand the accelerating expansion of the Universe. One approach to studying such models involves explicitly parameterizing the EoS of DE. The EoS parameter shows the relationship between pressure and energy density in DE. In the standard cosmological model, the EoS parameter is assumed to be constant throughout cosmic time with a value of $\omega{\Lambda}=-1$ for the cosmological constant. By permitting the EoS value to change over time, we can gain a better understanding of the fundamental physics that underlies DE. The Chevallier-Polarski-Linder (CPL) parametrization \cite{w-CPL1,w-CPL2,w-nCPL} is a useful two-parameter model that can capture deviations from a constant EoS value in the study of DE models. Beyond the CPL, there are more sophisticated parametrizations available, including the Jassal-Bagla-Padmanabhan (JBP) parametrization \cite{w-JBP}, the logarithmic parametrization \cite{w-LOG}, and the Barboza-Alcaniz (BA) parametrization \cite{w-BA}. These can also be employed to explore DE scenarios that go beyond the cosmological constant. Each parametrization has its own unique features and is capable of capturing different aspects of DE behavior. In addition, these parametrizations offer a flexible framework for testing different DE models and comparing them with observational data. Tab. \ref{tab1} provides a summary of the most commonly used parametrizations of EoS for DE in the literature.

Here, we consider a novel approach to parameterize the EoS for SFDE, which is expressed by a simple functional form,
\begin{equation}
\omega _{\phi }\left( z\right) =-1+\frac{2 \alpha  (1+z)^2}{3\left(\alpha  (1+z)^2+\beta\right)},
\label{EoS_phi}
\end{equation}%
where $\alpha $ and $\beta$ are constants. The set of equations is now complete/closed. The motivation for the proposed parameterization in Eq. (\ref{EoS_phi}) stems from its appealing features and potential to advance our understanding of SFDE and cosmological evolution. With only two parameters, $\alpha$ and $\beta$, the functional form offers simplicity and flexibility, making it efficient for cosmological analyses. Its physical interpretability allows $\alpha$ to govern the evolution of $\omega _{\phi }$, while $\beta$ serves as an offset term. It is clear that at $z=0$, which corresponds to the present epoch, the expression simplifies to $\omega_\phi(0)=-1+\frac{2\alpha}{3(\alpha+\beta)}$ and thus depends on the specific values of the parameters $\alpha$ and $\beta$. For instance, if $\beta=0$, then the EoS parameter at present is $-1/3$. On the other hand, the EoS parameter does tend to $\omega_\phi=-1/3$ at high redshift (i.e. $z\rightarrow \infty$). So, the proposed parametrization aligns with the well-established result that the universe is currently in a state of acceleration due to DE dominance. Because Eq. (\ref{EoS_phi}) describes the behavior of the EoS parameter for the SFDE component, the condition $\omega_\phi=-1/3$ at $z=0$ and $z\rightarrow \infty$ ensures that the model is consistent with the observed late-time cosmic acceleration. At very low redshift (i.e. $z\rightarrow -1$), the EoS parameter tends to $\omega_\phi=-1$, which indicates that the Universe would be dominated by a cosmological constant-like DE. Moreover, this parameterization can accommodate various cosmological scenarios, including early DE models (such as cosmic inflation) \cite{Liddle}, modified gravity theories, and other exotic cosmological models. Its ability to fit different cosmological data allows for investigations into a wide range of theoretical possibilities. The main features of this assumption are presented below in brief:
\begin{itemize}
    \item It is a time-varying parametrization, as the EoS depends on the redshift $z$.
    \item The EoS asymptotically approaches $\omega_\phi = -1$ as $z$ approaches -1, which is equivalent to the EoS of the cosmological constant.
    \item The parameter $\alpha$ controls the amplitude of the deviation from $\omega_\phi = -1$, while the parameter $\beta$ controls the redshift at which the deviation reaches its maximum.
    \item The EoS crosses the phantom divide line $\omega_\phi = -1$ at a redshift given by $z_c =-1- \sqrt{\frac{\beta}{\alpha}}$.
    \item For $\alpha > 0$ and $\beta > 0$, the EoS is always greater than $-1$, which corresponds to a quintessence-like behavior for DE.
    \item For $\alpha < 0$ and $\beta > -\alpha$, the EoS can cross the phantom divide line and become less than $-1$, which corresponds to a phantom-like behavior for DE.
\end{itemize}

\begin{widetext}
\begin{center}
\begin{table}[h]
\begin{tabular}{lcl}
\hline \hline
\multicolumn{1}{|l}{{\small EoS}} & 
\multicolumn{1}{|c}{\small Ref.} & \multicolumn{1}{|l|}{} \\ \hline \hline
\multicolumn{1}{|l}{${\small \omega(z)=\omega}_{0}{\small +\omega}_{1}{\small z}$} & 
\multicolumn{1}{|c}{{\small \cite{w-LIN1}, \cite{w-LIN2}}} & 
\multicolumn{1}{|l|}{\small Linear parametrization} \\ \hline
\multicolumn{1}{|l}{${\small \omega(z)=\omega}_{0}{\small +\omega}_{1}\frac{z}{\left(
1+z\right) ^{2}}$} & \multicolumn{1}{|c}{{\small \cite{w-JBP}}} & 
\multicolumn{1}{|l|}{\small JBP parametrization} \\ \hline
\multicolumn{1}{|l}{${\small \omega(z)=w}_{0}{\small +\omega}_{1}\frac{z}{1+z}$} & 
\multicolumn{1}{|c}{{\small \cite{w-CPL1}, \cite{w-CPL2}}} & 
\multicolumn{1}{|l|}{\small CPL parametrization} \\ \hline
\multicolumn{1}{|l}{${\small \omega(z)=\omega}_{0}{\small +\omega}_{1}\frac{z}{\sqrt{1+z^{2}%
}}$} & \multicolumn{1}{|c}{{\small \cite{w-sqrt}}} & \multicolumn{1}{|l|}%
{\small Square-root parametrization} \\ \hline
\multicolumn{1}{|l}{${\small \omega(z)=\omega}_{0}{\small +\omega}_{1}\frac{z(1+z)}{1+z^{2}}
$} & \multicolumn{1}{|c}{{\small \cite{w-BA}}} & \multicolumn{1}{|l|}{\small %
BA parametrization} \\ \hline
\multicolumn{1}{|l}{${\small \omega(z)=\omega}_{0}{\small +\omega}_{1}\frac{z}{1+z^{2}}$} & 
\multicolumn{1}{|c}{{\small \cite{w-FSSL}}} & \multicolumn{1}{|l|}{\small %
FSLL I parametrization} \\ \hline
\multicolumn{1}{|l}{${\small \omega(z)=\omega}_{0}{\small +\omega}_{1}\frac{z^{2}}{1+z^{2}}$%
} & \multicolumn{1}{|c}{{\small \cite{w-FSSL}}} & \multicolumn{1}{|l|}%
{\small FSLL II parametrization} \\ \hline
\multicolumn{1}{|l}{${\small \omega(z)=\omega}_{0}{\small +\omega}_{1}\sin {\small (z)}$} & 
\multicolumn{1}{|c}{{\small \cite{w-sin}}} & \multicolumn{1}{|l|}{\small Sine parametrization%
} \\ \hline
\multicolumn{1}{|l}{${\small \omega(z)=\omega}_{0}{\small +\omega}_{1}\ln {\small (1+z)}$}
& \multicolumn{1}{|c}{{\small \cite{w-LOG}}} & \multicolumn{1}{|l|}{\small %
Logarithmic parametrization} \\ \hline
\multicolumn{1}{|l}{${\small \omega(z)=w}_{0}{\small +\omega}_{1}\left( \frac{\ln (2+z)%
}{1+z}-\ln 2\right) $} & \multicolumn{1}{|c}{{\small \cite{w-MZ}}} & 
\multicolumn{1}{|l|}{\small MZ I parametrization} \\ \hline
\multicolumn{1}{|l}{${\small \omega(z)=\omega}_{0}{\small +\omega}_{1}\left( \frac{\sin
(1+z)}{1+z}-\sin 1\right) $} & \multicolumn{1}{|c}{{\small \cite{w-MZ}}} & 
\multicolumn{1}{|l|}{\small MZ II parametrization} \\ \hline
\multicolumn{1}{|l}{${\small \omega(z)=\omega}_{0}{\small +\omega}_{1}\left( \frac{z}{1+z}%
\right) ^{n}$} & \multicolumn{1}{|c}{{\small \cite{w-nCPL}}} & 
\multicolumn{1}{|l|}{\small Generalized CPL parametrization} \\ \hline
\multicolumn{1}{|l}{${\small \omega(z)=\omega}_{0}{\small +\omega}_{1}\frac{z}{\left(
1+z\right) ^{n}}$} & \multicolumn{1}{|c}{{\small \cite{w-nCPL}}} & 
\multicolumn{1}{|l|}{\small Generalized JBP parametrization} \\ \hline
\multicolumn{1}{|l}{${\small \omega(z)=\omega}_{0}{\small +\omega}_{1}\ln \left( {\small 1+%
\frac{z}{1+z}}\right) $} & \multicolumn{1}{|c}{{\small \cite{w-feng}}} & 
\multicolumn{1}{|l|}{\small Logarithmic parametrization} \\ \hline \hline
  \end{tabular}
     \caption{Parametrizations of the EoS for DE in different forms.}
 \label{tab1}
    \end{table}
\end{center}
\end{widetext}

By using Eqs. (\ref{rhom}), (\ref{t_z}), (\ref{EoS}), and (\ref{EoS_phi}), we can derive the following differential equation:
\begin{widetext}
\begin{equation}
\frac{-3 H^2(z)-2 (-(1+z)) H(z) \frac{\partial H}{\partial z}}{3 H^2(z)-3 H_{0}^2 \Omega_{m0} (1+z)^3}-\frac{2 \alpha  (1+z)^2}{3 \left(\beta +\alpha  (1+z)^2\right)}+1=0.
\label{HE}
\end{equation}
\end{widetext}

By solving Eq. (\ref{HE}), one can obtain the solution,
\begin{equation}
    H(z)=H_{0} \sqrt{\Omega_{m0} (1+z)^3+\frac{(1-\Omega_{m0})}{\alpha +\beta }\left(\alpha  (1+z)^2+\beta\right)},
    \label{Hz}
\end{equation}
where $H_0$ is the present value of the Hubble parameter (i.e. at $z=0$) and $\Omega _{m0}=\frac{\rho _{m0}}{3H_{0}^{2}}$ is the present value of the matter density parameter. This equation gives the Hubble parameter $H(z)$ as a function of the redshift $z$ for our EoS parametrization of DE. At high redshifts ($z >> 1$), the first term in the square root dominates, and $H(z)$ approaches $H_0 \sqrt{\Omega_{m0}z^3}$, while at low redshift ($z << 1$) and at the present time, the second term dominates, and $H(z)$ approaches $H_0$, which is the current expansion rate of the Universe. To reduce our model to $\Lambda$CDM, we need to set $\alpha=0$. This would make the Hubble parameter for our EoS parametrization reduce to:
\begin{equation}
    H(z) = H_0 \sqrt{\Omega_{m0}(1+z)^3 + \Omega_{\Lambda0}},
\end{equation}
which is exactly the same as the Hubble parameter for the $\Lambda$CDM model with the cosmological constant density parameter is $\Omega_{\Lambda0}=(1-\Omega_{m0})$.

The deceleration parameter $q$, which is defined as $(-\frac{\overset{..}{a}a}{\overset{.}{a}^{2}})$, can be expressed as a function of redshift z as follows:
\begin{equation}
q(z)=-1-\frac{\left( 1+z\right) }{H\left( z\right) }\frac{dH\left( z\right) }{%
dz}.  \label{q}
\end{equation}

Using Eqs. (\ref{Hz}) and (\ref{q}), the present model yields the following expression for the deceleration parameter,
\begin{equation}
q(z)=\frac{\alpha  \Omega_{m0} (1+z)^3+\beta  (\Omega_{m0} (z (z (3+z)+3)+3)-2)}{2 \left(\beta +\alpha  (1+z)^2 (1+\Omega_{m0} z)+\beta  \Omega_{m0} z (z (3+z)+3)\right)}.
\label{qz}
\end{equation}

One can investigate the evolution of $q(z)$ by using the constrained values of the parameters $\Omega_{m0}$, $\alpha$, and $\beta$ obtained from observations. Therefore, we can express the density $\rho_{\phi}$ and pressure $p_{\phi}$ as,
\begin{equation}
\rho_{\phi}(z)=-\frac{3 H_{0}^2 (\Omega_{m0}-1)}{\alpha +\beta }\left(\alpha  (1+z)^2+\beta\right),
\end{equation}
and
\begin{equation}
p_{\phi}(z)=\frac{H_{0}^2 (\Omega_{m0}-1)}{\alpha +\beta }\left(\alpha  (z+1)^2+3 \beta\right).
\end{equation}

Therefore, the total EoS parameter for our model can be expressed as,
\begin{widetext}
\begin{equation}
\omega_{T}(z)=\frac{(\Omega_{m0}-1) \left(\alpha  (1+z)^2+3 \beta\right)}{3 \left(\beta +\alpha  (1+z)^2 (\Omega_{m0} z+1)+\beta  \Omega_{m0} z (z (3+z)+3)\right)}.
\end{equation}

The density parameters for the matter field $\Omega _{m}$ and the scalar field $\Omega _{\phi }$ can be obtained in terms of redshift $z$ for this model as follows:
\begin{equation}
\Omega _{m}\left( z\right) =\frac{\rho _{m}}{3H^{2}}=\frac{\Omega_{m0} (1+z)^3 (\alpha +\beta )}{\beta +\alpha  (1+z)^2 (\Omega_{m0} z+1)+\beta  \Omega_{m0} z (z (3+z)+3)},
\end{equation}%
and%
\begin{equation}
\Omega _{\phi }\left( z\right) =\frac{\rho _{\phi }}{3H^{2}}=-\frac{(\Omega_{m0}-1) \left(\alpha  (1+z)^2+\beta\right)}{\beta +\alpha  (1+z)^2 (\Omega_{m0} z+1)+\beta  \Omega_{m0} z (z (3+z)+3)},
\end{equation}%
respectively.
\end{widetext}

Using Eqs. (\ref{F1}), (\ref{F2}), the potential energy (PE) $V\left( \phi \right) $ and kinetic energy (KE) $\frac{\overset{.}{\phi }^{2}}{2}$ of the scalar field can be expressed in terms of redshift as,%
\begin{equation}
V\left( \phi \right) =-\frac{H_{0}^2 (\Omega_{m0}-1)}{\alpha +\beta }\left(2 \alpha  (1+z)^2+3 \beta\right),
\label{V_z}
\end{equation}%
and%
\begin{equation}
\frac{\overset{.}{\phi }^{2}}{2}=-\frac{\alpha  H_{0}^2 (\Omega_{m0}-1)}{\alpha +\beta }(1+z)^2.
\end{equation}%
respectively. In addition, Eq. (\ref{V_z}) provides the functional form of the scalar field potential $V(\phi)$ which is dependent on various model parameters i.e. $H_{0}$, $\Omega_{m0}$, $\alpha$ and $\beta$. Therefore, the specific values of these parameters have a crucial role in determining the functional behavior of the cosmological parameters. One approach is to arbitrarily select the values of these parameters and investigate the corresponding dynamical cosmological parameters. The resulting values can then be compared with observational data. However, in this study, we adopt a different approach. We first utilize the available datasets to constrain the various model parameters. Then, with the best-fit values, we reconstruct the functional dependence of $V(\phi)$. This approach allows us to obtain a more realistic and observationally consistent model for the Universe.

\section{Methodology and Analysis of Observational Data}
\label{sec4}

In this study, we aim to fit our theoretical model to several recent observational datasets, namely the Cosmic Chronometer (CC) dataset, Baryon Acoustic Oscillations (BAO) dataset, and Type Ia Supernova (SNe) dataset. By combining these datasets, we obtain constraints on the model parameters. Subsequently, in the following sections, we attempt to reconstruct the functional form of $V(\phi)$. To accomplish this, we will describe the techniques used for analyzing each of these datasets in the following subsections.

To begin, we employ Bayesian statistical analysis and the \textit{emcee} python library \cite{Mackey/2013} to perform a Markov chain Monte Carlo (MCMC) simulation, which allows us to obtain the posterior probability distribution of the model parameters. The posterior probability distribution is a probability distribution that represents our degree of belief in the values of the model parameters after taking into account the available observational data. Once the MCMC simulation is completed, we analyze the resulting chain of parameter values to obtain the best-fit values and the uncertainties on the model parameters. The parameters space for our model can be described as,
\begin{equation}
    \theta_{s}=(H_{0}, \Omega_{m0}, \alpha, \beta)
\end{equation}

In addition, the best-fit values for the parameters can be obtained using the probability function $\mathcal{L} \propto exp(-\chi^2/2)$, where $\chi ^{2}$ is the pseudo-chi-squared function. The $\chi ^{2}$ function is a statistical tool that is commonly used in cosmology to obtain the parameters of a specific cosmological model that best suits observed data. The objective is to determine the values of $\theta_{s}$ that yield the minimum value of $\chi^{2}$. To achieve this, an MCMC approach is typically utilized to explore the parameter space and identify the regions that have the highest likelihood given the observational data.

\subsection{$CC$ $dataset$}

In this study, we consider Hubble parameter measurements obtained using the differential age (DA) method, also known as the Cosmic Chronometer (CC) datasets. Specifically, we use 31 data points obtained from Refs. \cite{Yu/2018,Moresco/2015,Sharov/2018}, which provide Hubble parameter observations at different redshifts, allowing us to constrain the Universe's expansion history and test different cosmological models. The $\chi ^{2}$ function can be expressed as follows:
\begin{equation}
\chi _{CC}^{2}=\sum_{i}\frac{\left[ H(\theta_{s} ,z_{i})-H_{obs}(z_{i})\right]
^{2}}{\sigma (z_{i})^{2}},
\end{equation}%
where $i$ ranges from 1 to 31. Here, $H(\theta_s, z_i)$ represents the theoretical value of the Hubble parameter at redshift $z_i$ for a specific set of cosmological parameters $\theta_s$, $H_{obs}(z_i)$ represents the measured value of the Hubble parameter at redshift $z_i$, and $\sigma (z_{i})$ represents the corresponding uncertainty of $H_{i}$. 

\subsection{$BAO$ $dataset$}

Baryon Acoustic Oscillations (BAO) are patterns in the large-scale structure of the Universe that arise from initial density perturbations in the baryon-photon plasma during the early Universe. These perturbations generated pressure waves, which moved through the plasma until recombination, at which point the universe becomes transparent to photons. At this point, the pressure waves left an imprint on the distribution of matter, which is still observable in the clustering of galaxies today. By analyzing the positions of the peaks in the CMB radiation power spectrum and in the distribution of galaxies on large angular scales, the characteristic scale of the BAO can be estimated. This scale serves as a standard ruler for cosmic distance measurements and is related to the sound horizon at recombination. To constrain the expansion history of the Universe and test different cosmological scenarios, we utilize the BAO dataset from various surveys, including the 6dFGS ( Six Degree Field Galaxy Survey), the SDSS (Sloan Digital Sky Survey), and the LOWZ samples of the BOSS (Baryon Oscillation Spectroscopic Survey) \cite{BAO1,BAO2,BAO3,BAO4,BAO5,BAO6}. These surveys have produced precise measurements of the positions of the BAO peaks in the galaxy clustering at different redshifts. The characteristic scale of BAO is determined by the sound horizon $r_s$ at the epoch of photon decoupling with redshift $z_{dec}$, and is related by the following equation:
\begin{equation}\label{4b}
r_{s}(z_{\ast })=\frac{c}{\sqrt{3}}\int_{0}^{\frac{1}{1+z_{\ast }}}\frac{da}{
a^{2}H(a)\sqrt{1+(3\Omega _{b0}/4\Omega _{\gamma 0})a}},
\end{equation}
where $\Omega _{b0}$ and $\Omega _{\gamma 0}$ indicate the present densities of baryons and photons respectively. This study uses the BAO dataset comprising six points for $d_{A}(z_{\ast })/D_{V}(z_{BAO})$ obtained from Refs. \cite{BAO1, BAO2, BAO3, BAO4, BAO5, BAO6}. Here, $z_{\ast }\approx 1091$ represents the redshift at the epoch of photon decoupling, and $d_{A}(z)=c\int_{0}^{z}\frac{dy}{H(y)},$ denotes the co-moving angular diameter distance, along with the dilation scale $D_{V}(z)=\left[d_{A}^{2}(z)cz/H(z)\right] ^{1/3}$. The BAO datasets are evaluated using the chi-square function presented in \cite{BAO6} as follows:
\begin{equation}\label{4e}
\chi _{BAO}^{2}=X^{T}C_{BAO}^{-1}X,,
\end{equation}

Here, $X$ is a quantity that varies depending on the particular survey being considered, while $C_{BAO}^{-1}$ is the inverse of the covariance matrix \cite{BAO6}.

\subsection{$SNe$ $dataset$}

Type Ia Supernovae (SNe) are a significant tool for studying the Universe's accelerating expansion and the nature of DE. These SNe are the result of a white dwarf star exploding in a binary system and have a unique light curve that makes them valuable \textit{standard candles} for estimating cosmic distances. By comparing their observed luminosity to their theoretical intrinsic luminosity, we can estimate their distance and chart the expansion history of the Universe. The Pantheon sample is a valuable dataset of SNe with 1048 data points covering a broad range of redshifts $0.01\leq z \leq2.26$, constructed from the PanSTARSS1 Medium Deep Survey, SDSS, SNLS, and numerous low-z, and HST samples. This dataset has been extensively calibrated to minimize systematic errors and increase distance estimation accuracy, making it an important resource for modern cosmology research \cite{Scolnic/2018,Chang/2019}. The $\chi ^{2}$ function for SNe datasets is expressed as the sum over all data points $i$ of the squared difference between the theoretical and observed distance modulus, weighted by the inverse covariance matrix $C_{SNe}^{-1}$, i.e.,
\begin{equation}
\chi _{SN}^{2}=\sum_{i,j=1}^{1048}\Delta \mu _{i}\left( C_{SN}^{-1}\right)
_{ij}\Delta \mu _{j},
\end{equation}%
where $\Delta \mu_{i}=\mu_{\rm th}-\mu_{\rm obs}$ is the difference between the theoretical and observed distance modulus, and $\mu = m_{B}-M_{B}$ represents the difference between the apparent magnitude $m_{B}$ and the absolute magnitude $M_{B}$. The nuisance parameters in the above equation were estimated using the BEAMS with Bias Corrections (BBC) approach \cite{Kessler/2017}. The theoretical value of the distance
modulus is calculated as,%
\begin{equation}
\mu _{th}(z)=5log_{10}\frac{d_{L}(z)}{1Mpc}+25,
\end{equation}%
\begin{equation}
d_{L}(z)=c(1+z)\int_{0}^{z}\frac{dy}{H(y,\theta_{s} )},
\end{equation}%
where $d_{L}(z)$ denotes the distance that takes into account the attenuation of light due to the expansion of the Universe, commonly known as the luminosity distance.

\subsection{$CC+BAO+SNe$ $datasets$}

The total joint $\chi_{total}^{2}$ function is utilized to combine the CC, BAO, and SNe samples, which is expressed as 
\begin{equation}
    \chi_{total}^{2}=\chi_{CC}^{2}+\chi_{BAO}^{2}+\chi_{SNe}^{2}.
\end{equation}

The goal is to obtain the model parameters by minimizing the $\chi ^{2}$ value, which is a common approach used in maximum likelihood analysis. The combined $CC+BAO+SNe$ datasets mentioned earlier are employed to estimate the best-fit values of the model parameters. This approach enables us to obtain a more comprehensive understanding of the nature of DE and the Universe's accelerating expansion. The combined datasets provide a more robust and precise constraint on the model parameters, allowing for a more reliable interpretation of the results. Generally, the combination of CC, BAO, and SNe datasets allows for more powerful cosmological analysis and a deeper insight into the fundamental properties of the Universe. In our MCMC analysis, we employed 100 walkers and 1000 steps to obtain the results. Fig. \ref{CC+BAO+SN} displays the marginalized constraints on the parameters space $\theta_{s}$ with $1-\sigma $ and $2-\sigma $ likelihood contours, and the numerical outcomes are listed in Tab. \ref{tab2}.

\begin{widetext}

\begin{figure}[h]
\centerline{\includegraphics[scale=0.8]{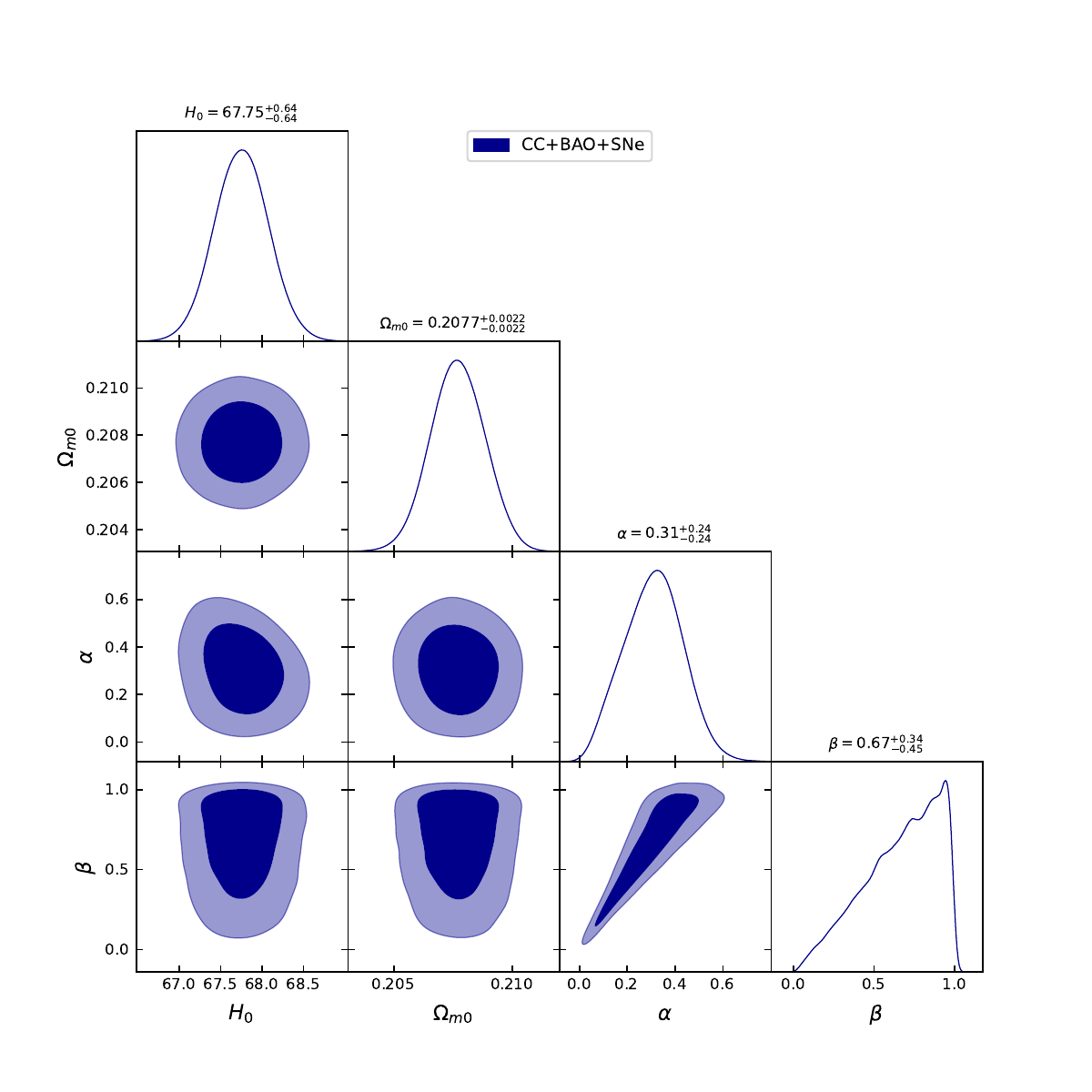}}
\caption{The plot depicts the $1-\sigma$ and $2-\sigma$ confidence level (CL) marginalized constraints on the parameters $H_{0}$, $\Omega_{m0}$, $\alpha$, and $\beta$ of our model using $CC+BAO+SNe$ datasets. The dark blue shaded zones indicate the $1-\sigma$ CL, while the light blue shaded zones represent the $2-\sigma$ CL. The constraint values for the parameters are also displayed at the $1-\sigma$ CL.}
\label{CC+BAO+SN}
\end{figure}

\begin{figure}[h]
\centerline{\includegraphics[scale=0.8]{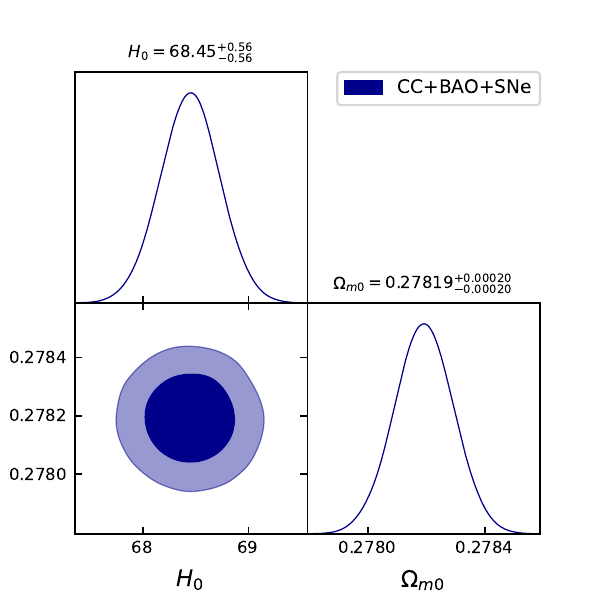}}
\caption{The plot presents the $1-\sigma$ and $2-\sigma$ confidence level (CL) marginalized constraints on the parameters $H_{0}$ and $\Omega_{m0}$ of the $\Lambda$CDM model using $CC+BAO+SNe$ datasets. The dark blue shaded zones indicate the $1-\sigma$ CL, while the light blue shaded zones represent the $2-\sigma$ CL. The constraint values for the parameters are also displayed at the $1-\sigma$ CL.}
\label{L_CC+BAO+SN}
\end{figure}

\begin{figure}[h]
\centerline{\includegraphics[scale=0.60]{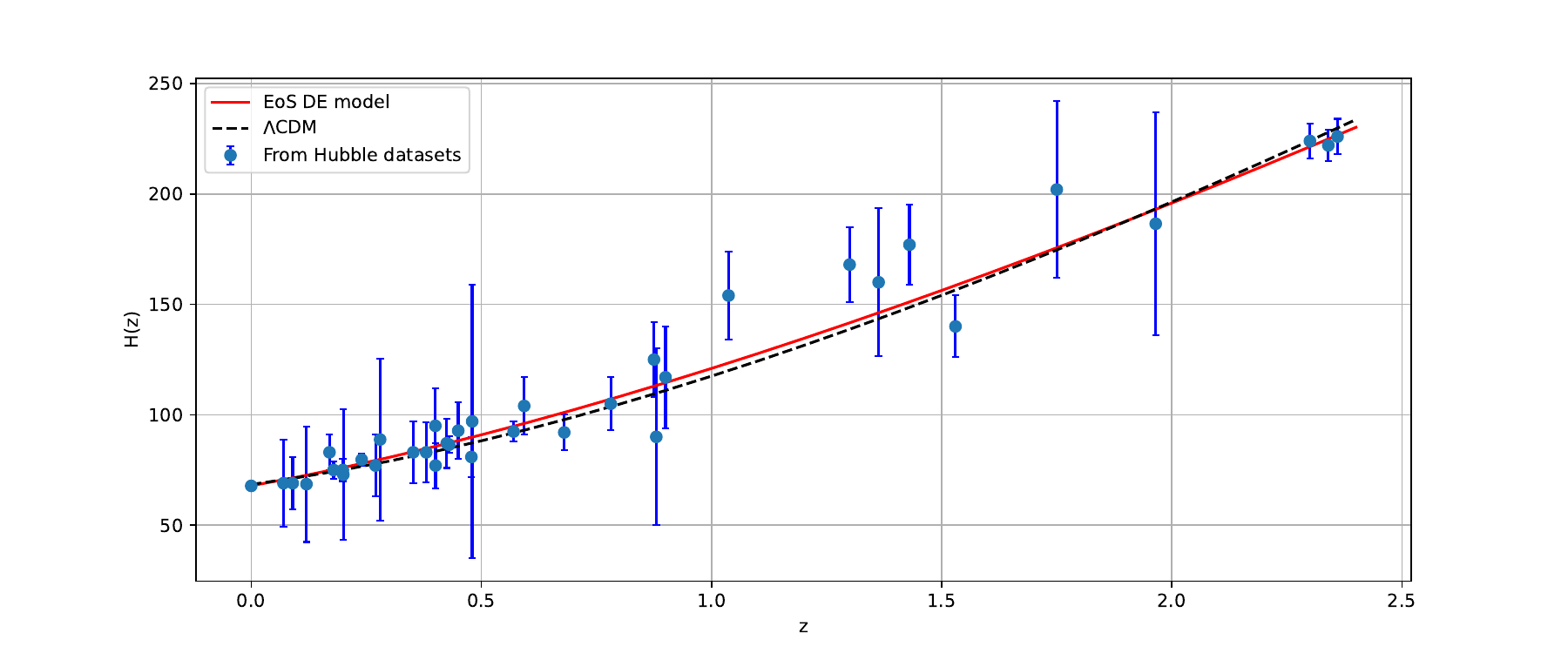}}
\caption{The plot presents a comparison between our model, represented by the red line, and the $\Lambda$CDM model, represented by the black dotted line, in terms of Hubble parameter $H(z)$ as a function of redshift $z$. The red line fits well with the 31 data points of the CC dataset, which are shown with their corresponding error bars.}
\label{ErrorHubble}
\end{figure}

\begin{figure}[h]
\centerline{\includegraphics[scale=0.60]{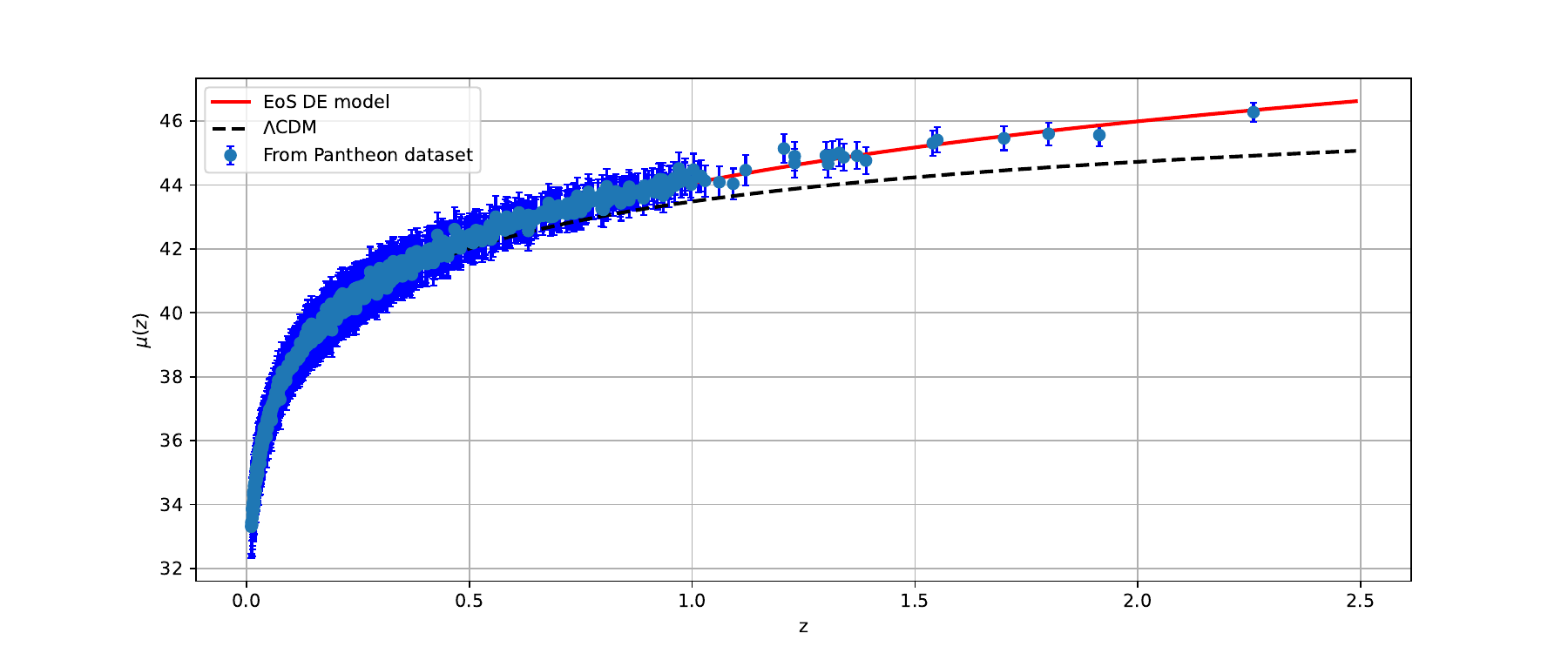}}
\caption{The plot presents a comparison between our model, represented by the red line, and the $\Lambda$CDM model, represented by the black dotted line, in terms of distance modulus $\mu(z)$ as a function of redshift $z$. The red line fits well with the 1048 data points of the SNe dataset, which are shown with their corresponding error bars.}
\label{ErrorSNe}
\end{figure}

\begin{table*}[!htb]
\begin{center}
\renewcommand{\arraystretch}{1.5}
\begin{tabular}{l c c c c c c c c c c}
\hline 
Parameters  & $H_{0}$ & $\Omega _{m0}$ & $\alpha$ & $\beta$ & $q_{0}$ & $z_{tr}$ & $\omega _{0}$\\
\hline
$Priors$ & $(60,80)$ & $(0,1)$  & $(0,1)$ & $(0,1)$ & $-$ & $-$ & $-$\\

$\Lambda CDM$   & $68.45_\pm0.56$  & $0.27819\pm0.00020$  & $-$ & $-$ & $-0.58\pm0.0003$ & $0.73\pm0.0006$ & $-1$\\

$SFDE$   & $67.75\pm0.64$  & $0.2077\pm0.0022$  & $0.31\pm0.24$ & $0.67_{-0.45}^{+0.34}$ & $-0.44_{-0.02}^{+0.03}$ & $0.73_{-0.01}^{+0.03}$ & $-0.79_{-0.01}^{+0.03}$\\

\hline 
Parameters  & $H_{0}$ & $\Omega _{m0}$ & $\omega_{0}$ & $\omega_{1}$ & $q_{0}$ & $z_{tr}$ & $-$\\
\hline
$CPL$   & $68.1\pm1.5$  & $0.290_{-0.067}^{+0.065}$  & $-1.05_{-0.24}^{+0.21}$ & $0.49_{-0.47}^{+0.49}$ & $-0.62_{-0.34}^{+0.31}$ & $0.61_{-0.01}^{+0.02}$ & $-$\\
\hline
\end{tabular}
\caption{The table presents the marginalized constraints on the model parameters for the $CC$+$BAO$+$SNe$ datasets, including the $\Lambda$CDM and CPL models, at a 68\% CL.}
\label{tab2}
\end{center}
\end{table*}
\end{widetext}

\section{Numerical Findings of Cosmological Parameters}
\label{sec5}
In this section, we will present and analyze the numerical results obtained from the statistical analysis. The cosmological parameters play a crucial role in understanding the evolution and structure of the Universe, and among them, the deceleration parameter is of particular significance as it characterizes the cosmic acceleration. We will focus on the marginalized constraints on the parameters included in the parameter space $\theta_{s}$, as obtained from the MCMC analysis. The combined $CC+BAO+SNe$ datasets were used to obtain two-dimensional likelihood contours, which include $1-\sigma $ and $2-\sigma $ errors corresponding to the $68\%$ and $95\%$ confidence levels. These likelihood contours for our model and $\Lambda$CDM are presented in Figs. \ref{CC+BAO+SN} and \ref{L_CC+BAO+SN}, respectively. The best-fit values for the model parameters are obtained as $H_{0}=67.75_{-0.64}^{+0.64}$, $\Omega_{m0}=0.2077_{-0.0022}^{+0.0022}$, $\alpha=0.31_{-0.24}^{+0.24}$, and $\beta=0.67_{-0.45}^{+0.34}$. On the other
hand, for the $\Lambda$CDM, the corresponding best-fit values are obtained as $H_{0}=68.45_{-0.56}^{+0.56}$, and $\Omega_{m0}=0.27819_{-0.00020}^{+0.00020}$ \cite{Mamon1,Mamon2}. It is important to note that the best-fit value of $\Omega_{m0}$ obtained in this study deviates substantially from the value reported by the Planck measurements \cite{Planck2020}. Our results indicate that the value of $\Omega_{m0}$ is significantly lower compared to the Planck value, with a value of nearly 30\%. This discrepancy may arise due to the use of different datasets, statistical methods, or theoretical models. Nonetheless, our findings still provide valuable insights into the constraints on $\Omega_{m0}$. In addition, we have obtained the numerical value of the present Hubble parameter, and our analysis shows that it is in good agreement with recent measurements from the Planck satellite, which reported a value of $H_0=67.4\pm0.5 km/s/Mpc$ \cite{Planck2020}. This finding is consistent with other studies that used similar methods to estimate the value of $H_0$, such as \cite{Chen1, Chen2, Aubourg, Capozziello}. Furthermore, our analysis has yielded positive values for both $\alpha$ and $\beta$, indicating a quintessence-like behavior for DE. Figs. \ref{ErrorHubble} and \ref{ErrorSNe} illustrate the plots of error bars for both the considered model and the standard cosmological model, $\Lambda$CDM. As can be observed in the figures, for both cases, our model exhibits close agreement with the observed data. Therefore, our result provides additional evidence supporting the current understanding of the cosmological parameters.

\begin{figure}[h]
\centerline{\includegraphics[scale=0.65]{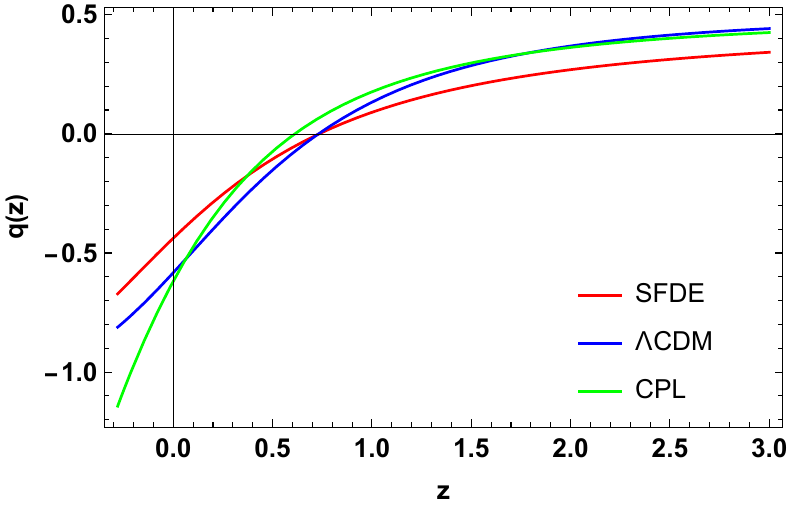}}
\caption{The plot presents the evolution of the deceleration parameter ($q$) vs. redshift ($z$), based on the values constrained from the $CC$+$BAO$+$SNe$ data-sets. Our model is compared to the standard $\Lambda$CDM and CPL models in the same plot.}
\label{F_q}
\end{figure}

Now, we present the evolution of various cosmological parameters of our cosmological model, such as deceleration, density, and EoS parameters, based on the constrained values of the model parameters. From Fig. \ref{F_q}, it is clear that the deceleration parameter $q$ in all cases, including that of $\Lambda$CDM and CPL models, smoothly transitions from a decelerating ($q>0$) to an accelerating phase ($q<0$) of the Universe for the $CC+BAO+SNe$ datasets. We also determined the observational constraints on the transition redshift $z_{tr}$ (i.e. at $q=0$), for the $CC+BAO+SN$ datasets and obtained their corresponding best-fit values. These values, along with the constraints from other parameterizations, are presented in Tab. \ref{tab2}. Our findings are in agreement with independent studies by several authors (refer to \cite{Farooq, Jesus, Garza}), which suggest that the Universe underwent a transition from decelerating to accelerating expansion at a redshift $z < 1$. However, it is noteworthy that the deceleration parameter evolution obtained from the joint analysis of $CC+BAO+SNe$ dataset is comparable to that of the $\Lambda$CDM model.

\begin{figure}[h]
\centerline{\includegraphics[scale=0.65]{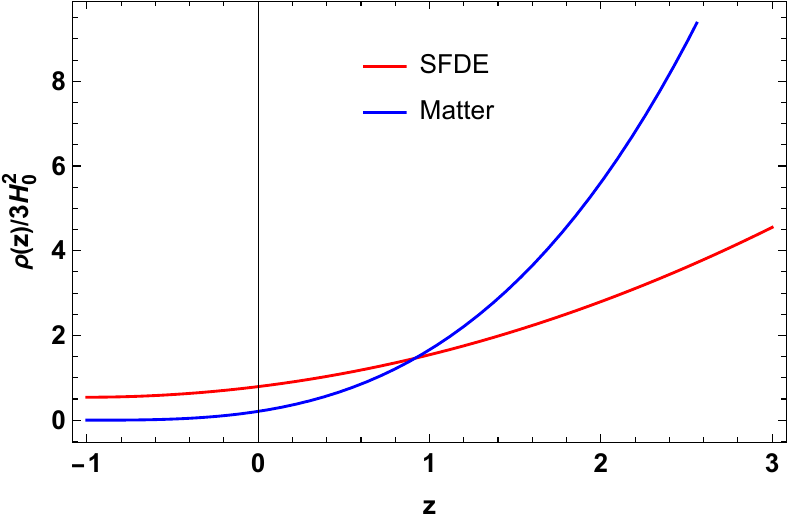}}
\caption{The plot presents the evolution of the densities of SFDE and matter ($\rho_{\phi}$ and $\rho_{m}$) vs. redshift ($z$), based on the values constrained from the $CC$+$BAO$+$SNe$ datasets.}
\label{F_rho}
\end{figure}

 Fig. \ref{F_rho} depicts the evolution of matter density and SFDE density as the Universe expands. As we can see from the figure, the densities of both matter and SFDE decrease as the Universe expands. In the late-time Universe, the matter density approaches zero, while the SFDE density approaches a minimum value. This behavior is a result of energy conservation in GR. Furthermore, the fact that the SFDE density tends to have a low value suggests that the Universe can also be able to expand at an accelerated rate in the future.

\begin{figure}[h]
\centerline{\includegraphics[scale=0.65]{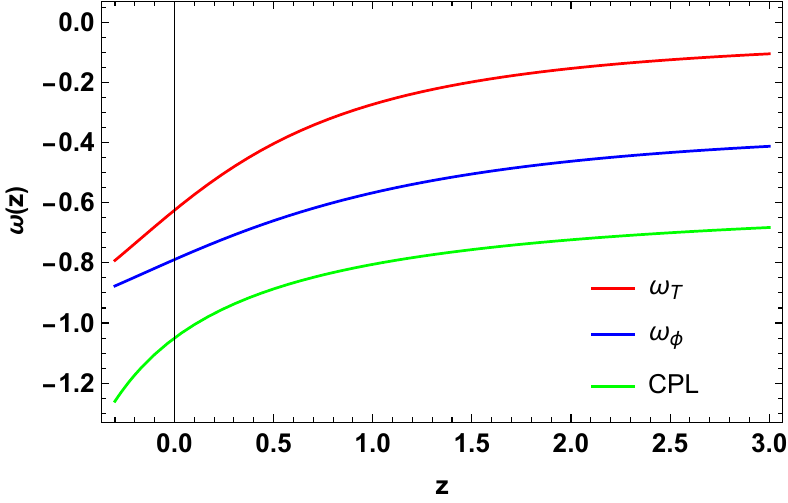}}
\caption{The plot presents the evolution of the EoS parameter ($\omega_{\phi}$ and $\omega_{eff}$) vs. redshift ($z$), based on the values constrained from the $CC$+$BAO$+$SNe$ datasets. Our model is compared to the CPL model in the same plot.}
\label{F_EoS}
\end{figure}

The EoS parameter is a useful tool for characterizing the behavior of the Universe in terms of its expansion rate. It is defined as the ratio of pressure $(p)$ to energy density $(\rho)$ of the cosmic fluid, and can take on different values depending on the nature of the cosmic fluid. For instance, for non-relativistic matter, such as dark matter, the EoS parameter is $\omega=0$, while for relativistic matter, such as radiation, it is $\omega=\frac{1}{3}$. The value of the EoS parameter can be used to categorize the Universe's decelerating and accelerating behavior. For a Universe with positive acceleration, there are three possible eras: the quintessence era, characterized by $-1<\omega<-\frac{1}{3}$; the phantom era, characterized by $\omega<-1$; and the cosmological constant era, characterized by $\omega=-1$. The plot in Fig. (\ref{F_EoS}) shows that both the EoS parameter for the SFDE and the total EoS parameter display an accelerating behavior. The total EoS parameter starts in the region where matter dominates and evolves through the quintessence era before settling at a constant value in the region where the cosmological constant dominates. Meanwhile, the EoS parameter for the SFDE exhibits quintessence behavior throughout the cosmic evolution and tends toward the cosmological constant in the future, which is consistent with the behavior of the total EoS parameter. For the CPL parameterization, we observe intriguing behavior in the $\omega_{CPL}$. At high redshifts, we find that $\omega_{CPL}$ lies in the quintessence region, indicating the dominance of a quintessence-like DE component. However, as we move towards relatively low redshifts, we observe a remarkable transition in the DE behavior. The EoS parameter enters the phantom regime. Interestingly, in the present epoch, our observations indicate that the $\omega_{CPL}$ is approaching values consistent with the $\Lambda$CDM model. The present value of the EoS parameter for the SFDE, which corresponds to the constrained values of the model parameters from $CC$+$BAO$+$SNe$ datasets, is $\omega_{0}=-0.79_{-0.01}^{+0.03}$, as reported in \cite{Hernandez, Zhang}.

\begin{figure}[h]
\centerline{\includegraphics[scale=0.65]{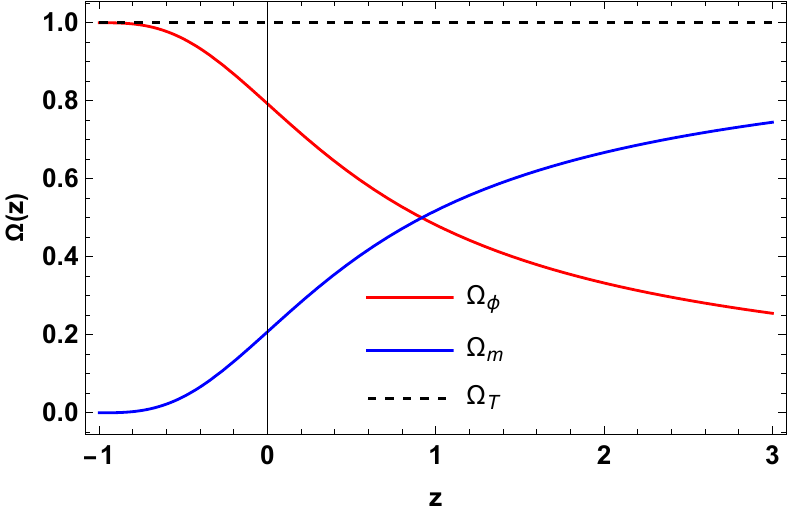}}
\caption{The plot presents the evolution of the density parameters ($\Omega_{\phi}$ and $\Omega_{m}$) vs. redshift ($z$), based on the values constrained from the $CC$+$BAO$+$SNe$ datasets.}
\label{F_Omega}
\end{figure}

In Fig. \ref{F_Omega}, the density parameter evolution for matter and scalar field are shown. Initially, the Universe is dominated by matter, while the SFDE density parameter is negligible. As the Universe expands, the matter density parameter decreases due to the increase in volume, while the SFDE density parameter becomes dominant, resulting in the acceleration of the Universe's expansion. 

\begin{figure}[h]
\centerline{\includegraphics[scale=0.65]{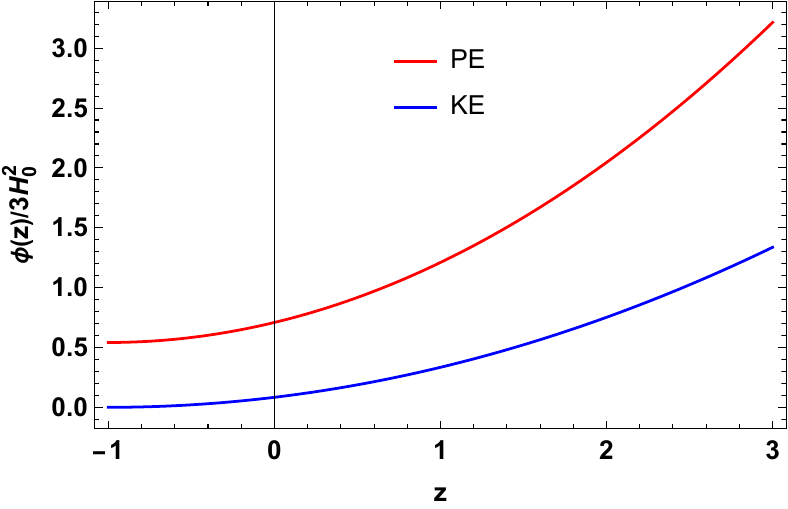}}
\caption{The plot presents the evolution of the potential energy (PE) and kinetic energy (KE) for the scalar field vs. redshift ($z$), based on the values constrained from the $CC$+$BAO$+$SNe$ datasets.}
\label{F_phi}
\end{figure}

In the context of DE, a scalar field is an enigmatic form of energy that permeates the Universe and is thought to be responsible for the accelerating expansion of the Universe. In addition, the kinetic and potential energies of the scalar field are important quantities in understanding its evolution. Fig. \ref{F_phi} illustrates the reconstructed behavior of the kinetic and potential energies of the scalar field over time. Based on the analysis in Fig. \ref{F_phi}, it can be observed that the potential term $V(\phi)$ changes at a much faster rate with redshift as compared to the kinetic term $\frac{\overset{.}{\phi }^{2}}{2}$ at the present epoch. This behavior is observed in several studies \cite{Pacif,Mamon1,Mamon2,S1,S2}.

\section{Analysis of Model's Geometrical Parameters}
\label{sec6}

\subsection{Statefinder diagnostics}
The concept of a cosmological constant $\Lambda$ is known to face significant issues such as the cosmological constant problem and the cosmic coincidence problem. In response, several dynamic models of DE have been proposed to overcome these shortcomings, as we have previously discussed in the introduction. However, it is essential to distinguish between these time-varying DE models to determine which one best fits observational data. To address this need, V. Sahni et al. \cite{Sahni, Alam} proposed the use of statefinder parameters, denoted as ($r$, $s$), as a new pair of geometrical parameters. These parameters provide a powerful tool for discriminating between different models of DE and have become a popular tool in modern cosmology. The statefinder parameters are defined as follows:
\begin{equation}
r=\frac{\overset{...}{a}}{aH^{3}}=2q^{2}+q-\frac{\overset{.}{q}}{H},  
\end{equation}%
\begin{equation}
s=\frac{\left( r-1\right) }{3\left( q-\frac{1}{2}\right) }.
\end{equation}

Using Eqs. (\ref{Hz}) and (\ref{qz}), we can express the statefinder parameters for our specific model as follows:
\begin{equation}
r(z)=\frac{\beta +\alpha  \Omega_{m0} (1+z)^3+\beta  \Omega_{m0} z (z (3+z)+3)}{\beta +\alpha  (1+z)^2 (\Omega_{m0} z+1)+\beta  \Omega_{m0} z (z (3+z)+3)},  
\end{equation}%
\begin{equation}
s(z)=\frac{2 \alpha  (1+z)^2}{3 \alpha  (1+z)^2+9 \beta}.
\end{equation}

The statefinder parameters can be used to distinguish between different DE models, as different models have different trajectories in the $r-s$ plane. The various values of the statefinder pair ($r$, $s$) correspond to different DE models, as described below:
\begin{itemize}
\item The $\Lambda$CDM scenario corresponds to $(r=1,s=0)$,

\item The SCDM (Standard Cold Dark Matter) scenario corresponds to $(r=1,s=1)$,

\item The HDE (Holographic DE) scenario corresponds to $(r=1,s=\frac{2}{3})$,

\item The CG (Chaplygin Gas) scenario corresponds to $(r>1,s<0)$,

\item The Quintessence scenario corresponds to $(r<1,s>0)$.\newline
\end{itemize}

\begin{figure}[h]
\centerline{\includegraphics[scale=0.60]{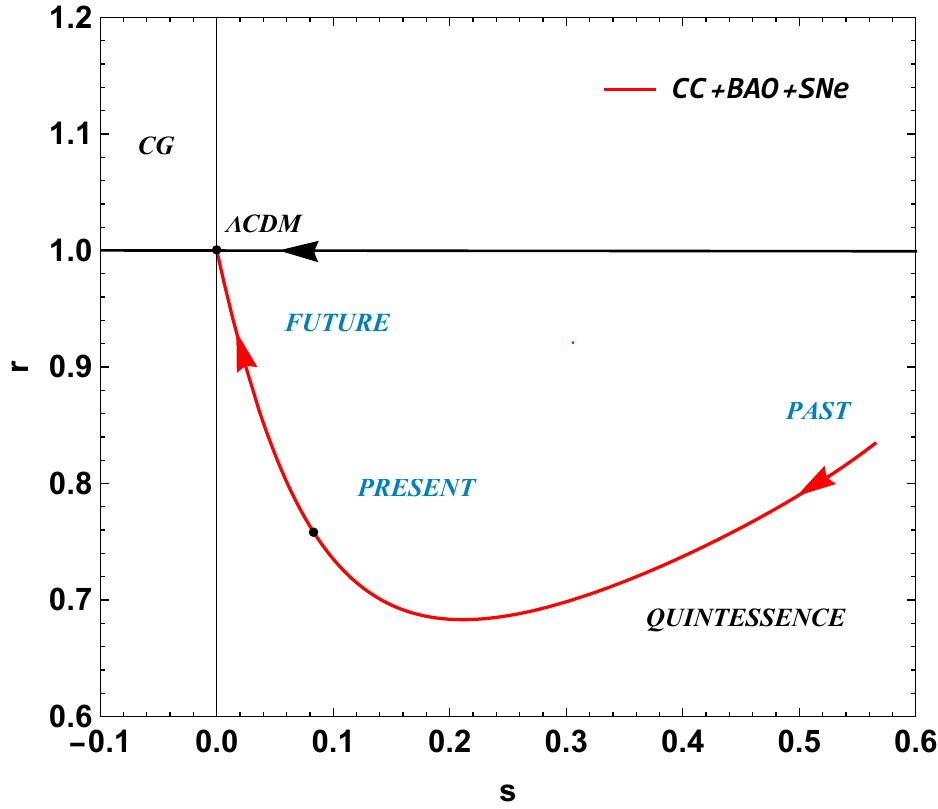}}
\caption{The plot presents the evolution of the $r-s$ plane, based on the values constrained from the $CC$+$BAO$+$SNe$ datasets with $-1\leq z\leq5$.}
\label{F_rs}
\end{figure}

\begin{figure}[h]
\centerline{\includegraphics[scale=0.59]{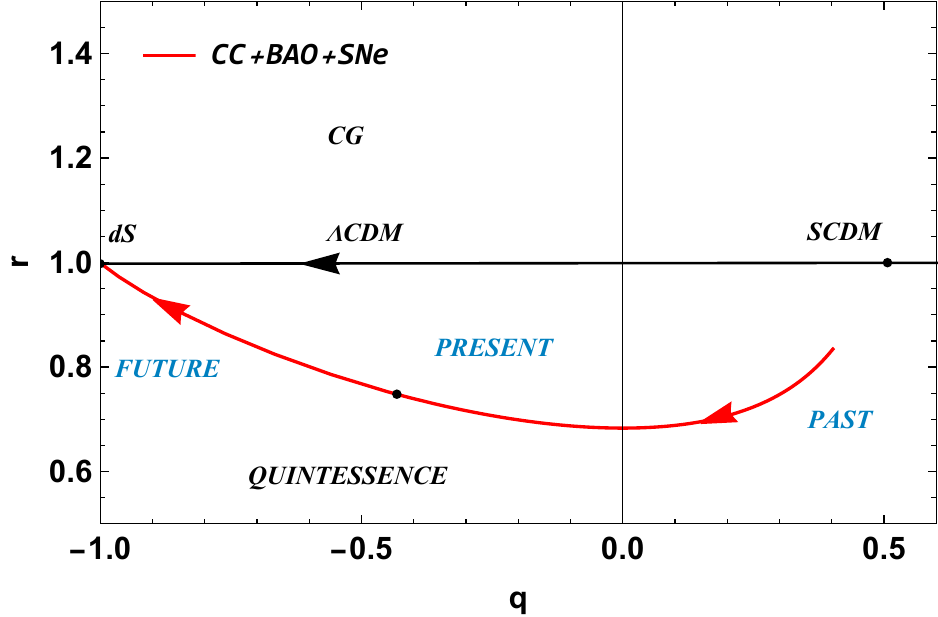}}
\caption{The plot presents the evolution of the $r-q$ plane, based on the values constrained from the $CC$+$BAO$+$SNe$ datasets with $-1\leq z\leq5$.}
\label{F_rq}
\end{figure}

Based on Fig. \ref{F_rs} in the $r-s$ plane, it can be observed that the model being studied initially has values of $r<1$ and $s>0$, which suggests that the scalar field behaves like quintessence and its energy density decreases as the Universe expands. This behavior is consistent with the current observational evidence of DE being responsible for the accelerating expansion of the Universe. However, over time, the model gradually approaches the $\Lambda$CDM model with $r=1$ and $s=0$. Similarly, Fig. \ref{F_rq} in the $r-q$ plane indicates that the current state of the Universe in the model is dominated by a quintessential fluid, but it is expected to evolve towards a de-Sitter (dS) phase ($r=1$ and $q=-1$) in which the Universe is dominated by a cosmological constant, leading to a constant rate of expansion. The behavior of the statefinder parameters in the model for the constrained values of the model parameters from the combined $CC$+$BAO$+$SNe$ datasets being studied is consistent with the behavior of the cosmological parameters discussed in the previous section. The model starts with a quintessence-like behavior, but eventually transitions to a $\Lambda$CDM-like behavior, indicating that the DE in the Universe changes over time. This is an important result, as it demonstrates that the statefinder parameters are useful tools for distinguishing between different DE models and understanding the evolution of the Universe.

\subsection{$Om(z)$ diagnostic}
The $Om(z)$ diagnostic provides another useful method to distinguish between different DE models in cosmology \cite{Sahni1}. It is a relatively simple diagnostic compared to the statefinder parameters, as it only involves the Hubble parameter, which is calculated by taking the first derivative of the cosmic scale factor. For a spatially flat Universe, the $Om(z)$ diagnostic is defined as follows:
\begin{equation}
Om(z) = \frac{\left(\frac{H(z)}{H_{0}}\right)^{2}-1}{(1+z)^{3}-1},
\label{Omz}
\end{equation}
where $H_{0}$ is the present-day Hubble parameter and $H(z)$ is the Hubble parameter at redshift $z$. 

Using Eqs. (\ref{Hz}) and (\ref{Omz}), we have
\begin{equation}
Om(z) = \frac{\beta \Omega_{m0}}{\alpha +\beta }+\frac{\alpha  \left(\Omega_{m0} (1+z)^2+z+2\right)}{(z (3+z)+3)(\alpha +\beta)}.
\label{Omzz}
\end{equation}

In particular, the slope of $Om(z)$ can provide information about the behavior of DE. A negative slope corresponds to quintessence behavior ($\omega>-1$), while a positive slope corresponds to phantom-type behavior ($\omega<-1$). A constant $Om(z)$ indicates the $\Lambda$CDM model ($\omega=-1$), where DE is described by a cosmological constant. From Eq. (\ref{Omzz}), it is clear that the $Om(z)$ diagnostic can be used to test the compatibility of the SFDE model with the $\Lambda$CDM model. It is known that the Om diagnostic should give a value of $\Omega_{m0}$ for the $\Lambda$CDM model. Therefore, to determine the values of $\alpha$ and $\beta$ at which the model reduces to $\Lambda$CDM, we set $\alpha=0$ and $\beta=1$ in the expression for $Om(z)$. In this case, the first term becomes $\Omega_{m0}$, and the second term becomes zero. Hence, $Om(z)$ reduces to $\Omega_{m0}$, which is the expected value for the $\Lambda$CDM model.

\begin{figure}[h]
\centerline{\includegraphics[scale=0.65]{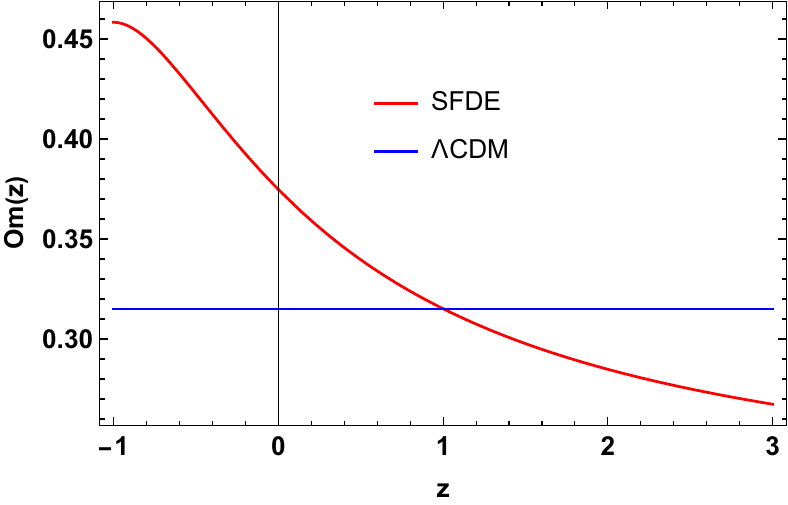}}
\caption{The plot presents the evolution of the $Om(z)$ diagnostic vs. redshift ($z$), based on the values constrained from the $CC$+$BAO$+$SNe$ datasets.}
\label{F_Om}
\end{figure}

In the present model, the behavior of $Om(z)$ can also be observed in Fig. \ref{F_Om} for the values constrained from the $CC$+$BAO$+$SNe$ datasets. The $Om(z)$ diagnostic shows a negative slope, indicating quintessence behavior. This is consistent with the previous results obtained for the statefinder parameters, which also suggest a transition from quintessence to $\Lambda$CDM-like behavior as the Universe evolves. Therefore, the $Om(z)$ diagnostic is a useful tool in analyzing the behavior of DE in the Universe and can provide additional insights into the evolution of cosmological models.

\section{Accretion and Evaporation Processes in the Presence of Scalar Field Dark Energy}
\label{sec7}

Let us begin by examining a static spherically symmetric metric, which is described by the following general equation:
\begin{equation}
ds^{2}=-A(r)dt^{2}+\frac{1}{B(r)}dr^{2}+r^{2}\left( d\theta ^{2}+\sin
^{2}\theta d\psi ^{2}\right),
\end{equation}
where $A(r)$ and $B(r)$ are functions of $r$ only and both are positive. This metric can be adjusted by choosing appropriate functions for $A(r)$ and $B(r)$ to represent a BH. We can assume that $M$ is the mass of the BH. For example, if we choose $A(r) = B(r) = 1- \frac{2M}{r}$, the resulting metric represents a Schwarzschild BH.

We can describe the process of matter flowing into a BH through accretion by considering a perfect fluid with an energy-momentum tensor given by Eq. (\ref{EMT}), which is appropriate for a spherically symmetric spacetime. Here, $\rho$ represents the energy density, $p$ represents the pressure, $u^\mu$ represents the four-velocity of the fluid, and $g^{\mu \nu}$ represents the metric tensor. The four-velocity of the fluid flow can be expressed as,
\begin{equation}
    u^{\mu}=\frac{dx^{\mu}}{ds}=(u^{0},u^{1},0,0),
\end{equation}
where $u^0$ and $u^1$ are only two non-zero components, which satisfy the condition $u^\mu u_\mu = -1$. This condition can be expressed in terms of the metric tensor components as $g_{00} u^0 u^0 + g_{11} u^1 u^1 = -1$. This implies that $u^0$ and $u^1$ are not independent, and the choice of one determines the other. Specifically, we can write $(u^0)^2 = \frac{(u^1)^2 + B}{AB}$, and since we are interested in the radial velocity of the fluid, we can let $u^1 = u$. Thus, we have $u_{0} = g_{00} u^0 = \sqrt{\frac{A}{B}} \sqrt{u^2 + B}$. We can use this expression for $u^0$ to determine the component $T^{1}_{0}$ of the energy-momentum tensor. From Eq. (\ref{EMT}), we can see that $T^{1}_{0}$ is proportional to $\rho + p$. Specifically, we obtain
\begin{equation}
    T^{1}_{0} = (\rho + p)u_{0} u.
\end{equation}

Here, we have used the expression for the determinant of the metric tensor $\sqrt{-g} = \sqrt{\frac{A}{B}} r^2 \sin\theta$. Also, we assume that the radial velocity $u$ is negative, corresponding to matter flowing into the BH.

In addition, we consider the fluid flow to consist of matter and SFDE. In order to obtain a proper DE accretion model for a static spherically symmetric BH, we need to generalize Michel's theory \cite{Michel}. Babichev et al. have already performed this generalization in the case of DE accretion onto a Schwarzschild BH \cite{Babichev1,Babichev2}. To properly treat the accretion process, the necessary fundamental equations are described by calculating the energy-momentum conservation equations and the particle number conservation equations. The energy-momentum conservation law is thus as follows:
\begin{align}
    T_{;\mu}^{\mu\nu}&=0,\\
     T_{;\mu}^{\mu\nu}&=\frac{1}{\sqrt{-g}}\bigg(\sqrt{-g}T^{\mu\nu}\bigg)_{,\mu}+\Gamma^\nu_{\alpha\mu}T^{\alpha\mu}=0,
\end{align}
where $\Gamma$ is the Christoffel symbol of the second kind and $(;)$ is the representation of covariant differentiation. By exploiting the structure of metric space-time, the above equation is clearly expressed as follows:
\begin{equation}
    T^{10}_{,1}+\frac{1}{\sqrt{-g}}T^{10}\bigg(\sqrt{-g}\bigg)_{,1}+2\Gamma_{01}^0\,T^{10}=0.
\end{equation}

We now have, after a few simplifications
\begin{equation}
    \frac{d}{dr}\left[r^2\,(\rho+p)\,u\frac{A(r)}{B(r)}\sqrt{u^2+B(r)}\right]=0,
\end{equation}
integrating the previous equation, we find
\begin{equation}
  r^2\,(\rho+p)\,u\frac{A(r)}{B(r)}\sqrt{u^2+B(r)}=c_1,
  \label{Eq48}
\end{equation}
where $c_1$ is the constant of integration. Using the law of conservation next to the four-velocity through $u_\mu\, T_{;\nu}^{\mu\nu}=0$, we obtain

\begin{widetext}
    \begin{equation}
        (p+\rho)_{,\nu}\,u_\mu u^\mu u^\nu+(p+\rho)\, u_{;\nu}^\mu u_\mu u^\nu+(p+\rho)u_\mu u^\mu u_{;\nu}^\nu+p_{,\nu}g^{\mu\nu}\, u_\mu+p\, u_\mu g_{;\nu}^{\mu\nu}=0.
    \end{equation}
\end{widetext}
Due to the fact that $g_{;\nu}^{\mu\nu} = 0$, and taking into account the normalization condition $(u^\mu u_\mu=-1)$, we obtain
\begin{equation}
    (p+\rho)\, u_{;\nu}^\nu+u^\nu\, \rho_{,\nu}=0,
\end{equation}
as long as $A_{;\,a}^b=\partial_a A^b+\Gamma_{ac}^b\, A^c$, we can thus find
\begin{equation}
    u^r\rho_{,r}+(p+\rho)\left[\Gamma_{0c}^0u^c+u_{,1}^1+\Gamma_{1c}^c u^c+\Gamma_{2c}^2 u^c+\Gamma_{3c}^3 u^c\right]=0.
    \label{eq51}
\end{equation}

After a few computations and considering the non-zero components of Eq. (\ref{eq51}), we arrive at dealing with a differential equation in the following compact form:
\begin{equation}
    \frac{\rho'}{p+\rho}+\frac{u'}{u}+\frac{A'}{2A}+\frac{B'}{2B}+\frac{2}{r}=0.
\end{equation}

In this case, a prime denotes differentiation with respect to $r$. By integrating the above equation, we get
\begin{equation}
    u\, r^2\sqrt{\frac{A(r)}{B(r)}}\, exp\bigg(\int\frac{d\rho}{p+\rho}\bigg)=-c_2.
\end{equation}

With $c_2$ is a constant of integration. As $u<0$, so in above equation $c_2 > 0$. The next examination finds out an expression resulting from the incorporation of the above equation with that of Eq. (\ref{Eq48}). So, the result is expressed as
\begin{equation}
    (p+\rho)\sqrt{u^2+B(r)}\,\sqrt{\frac{A(r)}{B(r)}}\,exp\bigg(-\int\frac{d\rho}{p+\rho}\bigg)=-\frac{c_1}{c_2}\equiv c_3,
\end{equation}
where $c_3$ is a constant of integration and has dependencies with $c_1$ and $c_2$. In addition, it is of interest to treat what is called the "mass flux", which is given by the following expression:
\begin{equation}
    (\rho\, u^\mu)_{;\mu}=\frac{1}{\sqrt{-g}}\bigg(\sqrt{-g}\rho\, u^\mu\bigg)_{,\mu}=0.
\end{equation}

Because our interest is on the equatorial plane, $\theta = \pi/2$, and the quantity $\sqrt{-g}\rho\, u^\mu$ is examined as a constant, we therefore have
\begin{equation}
    \rho\, u\sqrt{\frac{A(r)}{B(r)}}=\frac{c_4}{r^2},
    \label{Eq56}
\end{equation}
where $c_4$ is a constant. At the end of the dynamic equations discussion, we exploit Eq. (\ref{Eq48}) with (\ref{Eq56}) as the reason to obtain the following expression:
\begin{equation}
    \frac{(p+\rho)}{\rho}\sqrt{\frac{A(r)}{B(r)}}\,\sqrt{u^2+B(r)}=\frac{c_1}{c_4}\equiv c_5,
\end{equation}
where $c_5$ is an extra arbitrary constant.

To compute the rate of change of the BH mass $\dot{M}$, we integrate the flux of the fluid over the 2-dimensional surface of the BH, which is given by
\begin{equation}
   \dot{M} = -\int T^{1}_{0} \sqrt{-g}d\theta d\psi,
\end{equation}

Hence, the examination of the above integral with topological data on a 2-dimensional sphere can easily yield the expression of the rate of change of the mass of the BH as
\begin{equation}
   \overset{.}{M}=-4\pi r^2\, u(p+\rho)\sqrt{u^2+B(r)}\equiv -4\pi c_1.
\end{equation}

A closer examination shows $c_1=-c_2c_3$ and $c_3=\left( \rho _{\infty }+p\left( \rho _{\infty
}\right) \right)\sqrt{B(r_{\infty
})}$, the above-mentioned relation conducts to the following 
\begin{equation}
\overset{.}{M}=4\pi c_2 M^{2}\left( \rho _{\infty }+p\left( \rho _{\infty
}\right) \right)\sqrt{B(r_{\infty
})},
\end{equation}

Here, $\sqrt{B(r_{\infty
})}=1$, which results when the BH is a Schwarzschild case, $\rho _{\infty
}$ is the density at infinity, representing the matter or energy density in the far-reaching vicinity of a black hole, and its value depends on the specific astrophysical or cosmological scenario at hand. The derived result remains applicable for any EoS where the pressure is a function of the energy density i.e., $p=p(\rho)$ \cite{Babichev1,Babichev2}. Therefore, the rate of mass change for the accreting fluid surrounding the black hole can be expressed as
\begin{equation}
\overset{.}{M}_{acc}=4\pi c_2 M^{2}\left( \rho+p\right). 
\label{acc0}
\end{equation}

From Eq. (\ref{acc0}), it is clear that the rate at which a static BH with spherical symmetry changes mass due to the inflow of fluid is identical to the rate of change of mass for a Schwarzschild BH. In simpler terms, both types of BHs behave the same way when it comes to changes in their mass caused by fluid inflow. The above expression shows that the rate of change of mass for a static spherically symmetric BH does not depend on the functions $A(r)$ and $B(r)$. However, when fluid accretes outside the BH, the BH's mass function $M_{acc}$ becomes dynamic and can change over time. The rate of change of mass $\overset{.}{M}_{acc}$ is therefore dependent on time, and whether the BH's mass increases or decreases depends on the nature of the accreting fluid. If the sum of the density and pressure of the fluid (represented by $\rho + p$) is less than zero, which is the case for phantom DE accretion, the BH's mass decreases. On the other hand, if $\rho + p$ is greater than zero, which occurs with quintessence DE accretion, the BH's mass increases. It is worth noting that when the accretion fluid is only the cosmological constant, the mass of the BH in the accretion scenario remains constant throughout the time evolution.

For our SFDE model, we assume $\rho=\rho_{T}=\rho_{m}+\rho_{\phi}$ and $p=p_{T}=p_{\phi}$ \cite{Sarkar}. By using Eqs. (\ref{rho_phi}), (\ref{p_phi}), and (\ref{Hz}) in conjunction with Eq. (\ref{acc0}), we can obtain the function that expresses the mass of the BH resulting from accretion as a function of the redshift of the Universe as,
\begin{widetext}
\begin{equation}
{M}_{acc}(z)=\frac{\sqrt{\alpha +\beta }}{8 \pi  c_2 H_{0} \sqrt{\beta +\alpha  (1+z)^2 (\Omega_{m0} z+1)+\beta  \Omega_{m0} z (z (3+z)+3)}}.
\label{acc}
\end{equation}
\end{widetext}

\begin{figure}[h]
\centerline{\includegraphics[scale=0.65]{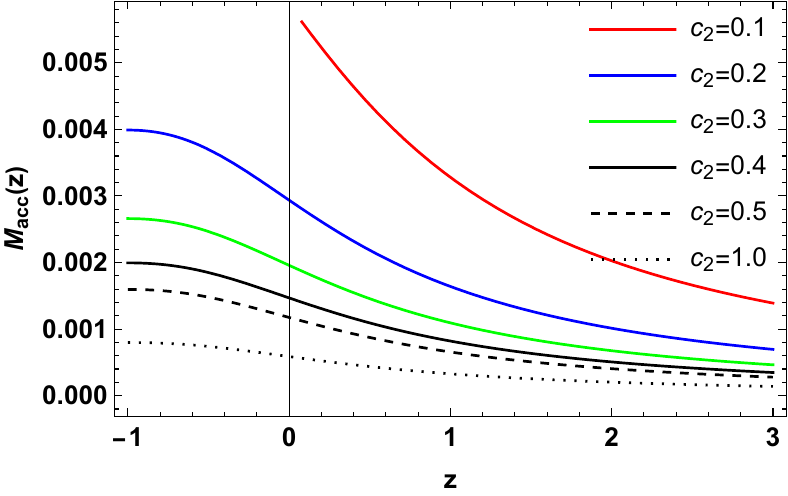}}
\caption{The plot presents the evolution of the BH mass $M_{acc}$ due to the accretion process vs. redshift ($z$), based on the values constrained from the $CC$+$BAO$+$SNe$ datasets.}
\label{F_M}
\end{figure}

Here, we have selected the constant of integration to be zero. From Fig. \ref{F_M}, we can observe that BHs within this scenario experience an increase in mass for some values of $c_2$, as expected (since the SFDE model behaves like a quintessence). However, this growth rate eventually stops once SFDE becomes the dominant component. For very long periods of cosmic time, the mass of the BH reaches a maximum value and no longer increases.

In addition to the accretion process, it is also possible to consider the evaporation of a BH through the emission of Hawking radiation. Also, Hawking radiation is a process through which BHs emit particles and lose mass over time i.e. $T_{H} = \frac{1}{M}$. This process is a result of quantum effects near the event horizon, where pairs of particles are created, with one member of the pair falling into the BH and the other escaping as radiation. As a result of this process, the mass of the BH decreases, and it eventually evaporates completely, leaving behind only radiation. The rate of evaporation depends on the mass of the BH \cite{Thorne,Haw1,Bek}, with smaller BHs evaporating faster than larger ones. The time it takes for a BH to evaporate completely is on the order of $10^{67}$ years for a BH with $M\sim M_{sun}$, and much longer for larger BHs \cite{Haw2}. In this case, the rate at which the BH mass changes due to evaporation can be expressed as \cite{Cline},
\begin{equation}
\overset{.}{M}_{eva}=-\frac{D}{M^{2}}.
\label{eva}
\end{equation}

Here, the value of the constant $D$ in Eq. (\ref{eva}) depends on the specific model being considered, and it is typically a positive quantity. The rate of change of the BH's mass is influenced by both the accretion of matter onto the BH and the evaporation of the BH itself through Hawking radiation. Now, this rate can be expressed as:
\begin{equation}
\overset{.}{M}=\overset{.}{M}_{acc}+\overset{.}{M}_{eva}=4\pi c_2M^{2}\left(
\rho+p\right) -\frac{D}{M^{2}}.
\end{equation}

These processes have different dependencies on the nature of the accreting or evaporating matter. In the case of accretion, the rate of change in the BH's mass is dependent on the nature of the accreting fluid. This means that the mass change will be different depending on whether the accreting matter is a normal fluid, a quintessence type DE fluid, or a phantom DE fluid. However, in the case of evaporation, the rate of change in the BH's mass is independent of the nature of the evaporating matter, since this is an internal process. When both accretion and evaporation are considered, $\overset{.}{M}>0$ for $M^{4}>\frac{D}{4\pi c_2\left( \rho _{T}+p_{T}\right) }$ and $\overset{.}{M}<0$ for $M^{4}<\frac{D}{4\pi c_2\left( \rho _{T}+p_{T}\right) }$ for a normal fluid and a
quintessence type DE. However, the BH mass always decreases $(\overset{.}{M}<0)$ in the presence of phantom DE. Therefore, evaporation can contribute to the reduction in the mass of the BH, but there are certain limitations on the minimum mass value that the BH can have.

\section{Concluding Remarks}
\label{sec8}
The search for an acceptable DE model that can accurately describe the evolution of the Universe has led to several proposals in cosmology. Among these proposals, dynamical models of DE have gained significant attention in the literature. These models provide a more comprehensive framework to study the evolution history of the Universe, taking into account both early-time and late-time scenarios. By exploring the dynamics of DE, it is possible to overcome some of the issues that arise in the cosmological constant model, such as the cosmic coincidence problem and the fine-tuning problem. Therefore, in this study, we have explored the behavior of a DE model with a scalar field in a spatially flat FLRW Universe. For this purpose, we have considered a novel approach to parameterize the EoS for SFDE, and from this, we have obtained analytical solutions for various cosmological parameters. The EoS parametrization used in our study is given by Eq. (\ref{EoS_phi}), which yields $\omega_\phi(0)=-1+\frac{2\alpha}{3(\alpha+\beta)}$ at the present epoch, and thus its value depends on the specific values chosen for the parameters $\alpha$ and $\beta$. Then, we employed the statistical MCMC method with the Bayesian technique to determine the constraint values for the parameters. We analyzed the outcomes for three distinct observational datasets, namely the CC, BAO, and SNe datasets. Furthermore, our analysis has yielded positive values for both $\alpha$ and $\beta$ as $\alpha=0.31_{-0.24}^{+0.24}$, and $\beta=0.67_{-0.45}^{+0.34}$, indicating a quintessence-like behavior for DE i.e. $\omega_{0}=-0.79_{-0.01}^{+0.03}$. The current value of the Hubble constant, denoted as $H_0$, is a crucial cosmological parameter that determines the present rate of cosmic expansion. Our analysis has yielded a value of $H_{0}=67.75_{-0.59}^{+0.59}km/s/Mpc$ \cite{Chen1, Chen2, Aubourg, Capozziello}, which is consistent with the most recent measurements from various cosmological probes \cite{Planck2020}. The precise determination of $H_0$ is essential for understanding the age, size, and fate of the Universe, as well as for testing different cosmological models. The matter density parameter also known as the fractional matter density describes the ratio of the matter density of the Universe to the critical density required for a flat Universe i.e. $\rho_{c}=3H^2$. In our analysis, we found the present value of the matter parameter to be $\Omega_{m0}=0.2077_{-0.0022}^{+0.0022}$ (Fig. \ref{F_Omega}), indicating that matter contributes to only $20.77\%$ of the total energy density of the Universe \cite{Mamon1,Mamon2}. This result is consistent with other observational studies and provides insights into the overall composition and evolution of the Universe.

Moreover, we have examined the evolution of the deceleration parameter, matter density, and SFDE density for the constrained values of the model parameters. The evolution of the deceleration parameter depicted in Fig. \ref{F_q} suggests that the Universe has transitioned from a decelerating phase to an accelerating phase recently, while both energy density displayed in Fig. \ref{F_rho} shows a positive behavior, as expected. The transition redshift, which corresponds to the values of the model parameters constrained by the $CC+BAO+SNe$ datasets is $z_{tr}=0.73_{-0.01}^{+0.03}$. Further, the present value of the deceleration parameter is $q_{0}-0.44_{-0.02}^{+0.03}$ for the $CC+BAO+SNe$ datasets \cite{Farooq, Jesus, Garza}. Furthermore, we investigated the reconstructed behavior of kinetic energy and potential energy of the scalar field. As shown in Fig. \ref{F_phi}, the potential energy of the scalar field is greater than its kinetic energy, i.e., $V\left( \phi \right)>\frac{\overset{.}{\phi }^{2}}{2}$. From Eq. (\ref{p_phi}), this implies that the SFDE has negative pressure $p_{\phi}<0$, and the potential energy $V\left( \phi \right)$ is the driving force behind the accelerated expansion of the Universe in the derived model.

In addition to the analysis of the SFDE model, we also studied the geometrical parameters of the model in Sec. \ref{sec6}. The behavior of the statefinder and $Om(z)$ diagnostics shown in Figs. \ref{F_rq}, \ref{F_rs} and \ref{F_Om}, respectively, further supported the quintessence-like behavior of DE in our derived model. The statefinder diagnostics plot revealed that our model is consistent with the values of $r<1$ and $s>0$, while the $Om(z)$ diagnostic plot suggested a deviation from the $\Lambda$CDM model with a negative slope. These results provide further insight into the validity and behavior of our derived model. In conclusion, the scalar field model of DE studied in this work is a viable alternative to the $\Lambda$CDM model, and its behavior is consistent with current observations. Future work could investigate the behavior of this model in more detail, including its impact on other cosmological parameters and the possibility of distinguishing it from other DE models using more advanced observational techniques.

Finally, we have analyzed the accretion behavior as well as the evaporation process of the most general static spherically
symmetric BH metric. The first phenomenon is simply a process for which the matter flow onto a BH surrounded a perfect fluid by accretion.  It is worth noting that the variation of  mass during the  accretion phenomenon is a nice indicator for having a suitable summary of that process. For that reason, we have taken the computation in terms of the mass accretion rate variation.
Some results regarding the accretion process have revealed that this process experiences a growth in mass for certain values of $c_2$, and during very long periods of cosmic time, the mass of the BH attains a maximum value and does not increase anymore. The second process is taken into account for such a quantum effect on the event horizon of a BH, namely, BH evaporation. In our study, a given scenario involves the BH's mass decreasing until it completely evaporates, leaving behind only radiation. The rate of evaporation varies with the mass of the BH, with smaller BHs evaporating faster than larger ones. Roughly speaking, the rate of change of the BH’s mass is influenced by both the accretion of matter onto the BH and the evaporation of the BH itself through Hawking radiation.

\section*{Acknowledgments}
The authors extend their appreciation to the Deanship of Scientific Research, Imam Mohammad Ibn Saud Islamic University (IMSIU), Saudi Arabia, for funding this research work through Grant No. (221412042).

\textbf{Data availability} This article does not introduce any new data.


\begin{thebibliography}{99}

\bibitem{DDE} E. J. Copeland, M. Sami, S. Tsujikawa, \textit{Int. J. Mod. Phys. D} \textbf{15}, 1753-1935 (2006).

\bibitem{Riess} A.G. Riess et al., \textit{Astron. J.} \textbf{116}, 1009
(1998).

\bibitem{Perlmutter} S. Perlmutter et al., \textit{Astrophys. J.} \textbf{517%
}, 565 (1999).

\bibitem{D.J.} D.J. Eisenstein et al., \textit{Astrophys. J.} \textbf{633},
560 (2005).

\bibitem{W.J.} W.J. Percival at el., \textit{Mon. Not. R. Astron. Soc.} 
\textbf{401}, 2148 (2010).

\bibitem{R.R.} R.R. Caldwell, M. Doran, \textit{Phys. Rev. D} \textbf{69},
103517 (2004).

\bibitem{Z.Y.} Z.Y. Huang et al., \textit{J. Cosm. Astrop. Phys.} \textbf{%
0605}, 013 (2006).

\bibitem{T.Koivisto} T. Koivisto, D.F. Mota, \textit{Phys. Rev. D} \textbf{73%
}, 083502 (2006).

\bibitem{S.F.} S.F. Daniel, \textit{Phys. Rev. D} \textbf{77}, 103513 (2008).

\bibitem{Planck2020} Planck Collaboration, \textit{Astron. Astrophys.} \textbf{641},
A6 (2020).

\bibitem{Sahni0} V. Sahnu and A. Starobinsky, \textit{Int. J. Mod. Phys. D} \textbf{09},
373-443 (2000).

\bibitem{dalal/2001} N. Dalal et al., \textit{Phys. Rev. Lett.} \textbf{87},
141302 (2001).

\bibitem{weinberg/1989} S. Weinberg, \textit{Rev. Mod. Phys.} \textbf{61}, 1
(1989).

\bibitem{HT0} A. G. Riess et al., \textit{Astrophys. J.} \textbf{876}, 85
(2019).

\bibitem{HT1} W. L. Freedman et al., \textit{Astrophys. J.} \textbf{882}, 34
(2019).

\bibitem{HT2} L. Verde, T. Treu, A. G. Riess, \textit{Nat. Astron} \textbf{3}, 891-895
(2019).

\bibitem{HT3} A. G. Riess et al., \textit{Astrophys. J.} \textbf{876}, 3
(2019).

\bibitem{Vagnozzi} S. Vagnozzi, \textit{Phys. Rev. D} \textbf{102}, 023518
(2020).

\bibitem{fR1} S. Appleby and R. Battye, \textit{Phys. Lett. B} \textbf{654}, 7-12 (2007).

\bibitem{fR2} L. Amendola et al., \textit{Phys. Rev. D} \textbf{75}, 083504 (2007).

\bibitem{fT1} M.R. Setare and N. Mohammadipour, \textit{J. Cosm. Astrop.Phys.} \textbf{01}, 015 (2013).

\bibitem{fT2} M. Koussour and M. Bennai, \textit{Class. Quantum Grav.} \textbf{39} 105001 (2022).

\bibitem{fQ1} J. B. Jim\'{e}nez et al., \textit{Phys. Rev. D} \textbf{98}, 044048 (2018).

\bibitem{fQ2} J.B. Jim\'{e}nez et al., \textit{Phys. Rev. D} \textbf{101}, 103507 (2020).

\bibitem{fQ3} M. Koussour et al., \textit{Phys. Dark Univ.} \textbf{36}, 101051 (2022).

\bibitem{fQ4} M. Koussour et al., \textit{J. High Energy Phys.} \textbf{37}, 15-24 (2023).

\bibitem{fQ5} M. Koussour and M. Bennai, \textit{Chin. J. Phys.} \textbf{79}, 339-347 (2022).

\bibitem{fQ6} M. Koussour et al.,{\ }\textit{Ann. Phys.} \textbf{445}, 169092 (2022).

\bibitem{fQ7} M. Koussour et al., \textit{J. High Energy Astrophys, } \textbf{35}, 43-51 (2022).

\bibitem{SF0} A. R. Liddle and R. J. Scherrer, \textit{Phys. Rev. D} \textbf{59}, 023509 (1999).

\bibitem{SF1} S. Dodleson, M. Kaplinghat and E. Stewart, \textit{Phys. Rev. Lett} \textbf{85}, 5276 (2000).

\bibitem{SF2} P. J. Steinhardt et al, \textit{Phys. Rev. D} \textbf{85}, 123504 (1999).

\bibitem{SF3} V. B. Johri, \textit{Class.Quant.Grav.} \textbf{19}, 5959 (2002).

\bibitem{SF4} V. B. Johri, \textit{Phys. Rev. D} \textbf{63}, 103504 (2001).

\bibitem{Ratra} B. Ratra and P. J. E. Peebles, \textit{Phys. Rev. D} \textbf{37}, 3406 (1988).

\bibitem{Peebles} P. J. E. Peebles and B. Ratra, \textit{Astrophys. J.} \textbf{325}, L17-L20 (1988).

\bibitem{Wetterich} C. Wetterich, \textit{Astron. Astrophys.} \textbf{301}, 321–328 (1995).

\bibitem{Carroll} S. M. Carroll, \textit{Phys. Rev. Lett.} \textbf{81}, 3067(1998).

\bibitem{Y.Fujii} Y. Fujii, \textit{Phys. Rev. D} \textbf{26}, 2580(1982).

\bibitem{Phantom1} S. Nojiri, S. D. Odintsov, and S. Tsujikawa, \textit{Phys. Rev. D} \textbf{71}, 063004 (2005).

\bibitem{Phantom2} R. R. Caldwell, M. Kamionkowski, and N. N. Weinberg, \textit{Phys. Rev. Lett.} \textbf{91}, 071301 (2003).

\bibitem{T.Chiba} T. Chiba et al., \textit{Phys. Rev. D} \textbf{62}, 023511(2000).

\bibitem{C.Arm.} C. Armendariz-Picon et al., \textit{Phys. Rev. Lett.} \textbf{85}, 4438(2000).

\bibitem{Kamenshchik} A. Y. Kamenshchik et al., \textit{Phys. Lett. B} 
\textbf{511}, 265 (2001).

\bibitem{M.C.} M. C. Bento et al., \textit{Phys. Rev. D} \textbf{66}, 043507(2002).

\bibitem{A.Y.} A. Y. Kamenshchik et al., \textit{Phys. Lett. B} \textbf{511}, 265(2001).

\bibitem{tachyon} J. S. Bagla, et al., \textit{Phys. Rev. D} \textbf{67},  063504 (2003).

\bibitem{Singh} J. K. Singh and R. Nagpal, \textit{Eur. Phys. J. C} \textbf{80}, 4 (2020).

\bibitem{Pacif} S. K. J. Pacif, S. Arora, and P. K. Sahoo, \textit{Phys. Dark Universe} \textbf{32}, 100804 (2021).

\bibitem{Bairagi} M. Bairagi, \textit{Phys. Dark Universe} \textbf{39}, 101158 (2023).

\bibitem{Debnath} U. Debnath and K. Bamba, \textit{Eur. Phys. J. C} \textbf{ 79}, 1-14 (2019).

\bibitem{Kar} A. Kar, S. Sadhukhan, and U. Debnath, \textit{Mod. Phys. Lett. A} \textbf{37}, 2250183 (2022).

\bibitem{Sharma} U. K. Sharma, M. Kumar, and G. Varshney, \textit{Universe} \textbf{8}, 12 (2022).

\bibitem{MK} M. Koussour, \textit{arXiv preprint} arXiv:2304.05807 (2023).

\bibitem{Cunha} J. V. Cunha and J. A. S. Lima, \textit{Mon. Not. R. Astr. Soc.}, 390, 210 (2008).

\bibitem{Mortsell} E. Mortsell and C. Clarkson, \textit{J. Cosm. Astropar. Phys.}, 2009, 01 (2009).

\bibitem{Pacif1} S. K. J. Pacif et al., \textit{Int. J. Geom. Meth. Mod. Phys.}, \textbf{14}, 7 (2017).
\bibitem{1}
S. K. Chakrabarti,  \textit{Physics Reports} {\bf 266} (1996) 229 – 390 .
\bibitem{2}
H. Bondi and F. Hoyle, \textit{Mon. Not. R. Astron. Soc} {\bf 104}, 273 (1944).
\bibitem{3}
F. C. Michel, \textit{Astrophys. Space Sci.} {\bf 15}, 153 (1972).
\bibitem{4}
M. Jamil, M. A. Rashid, and A. Qadir, \textit{Eur.Phys.J. C} {\bf 58}, 325 (2008).
\bibitem{5}
M. Sharif and G. Abbas, \textit{Chin. Phys. Lett} {\bf 28} (2011) 090402.
\bibitem{Barrow1} J.D. Barrow, \textit{Nucl. Phys. B}, \textbf{310}, 743 (1988).

\bibitem{Barrow2} J.D. Barrow, \textit{Phys. Lett. B}, \textbf{235}, 40 (1990).

\bibitem{w-CPL1} M. Chevallier and D. Polarski, Int. J. Mod. Phys. D 10 213
(2001).

\bibitem{w-CPL2} E. V. Linder, Phys. Rev. Lett. 90 091301 (2003).

\bibitem{w-nCPL} Dao-Jun Liu et al., Mon. Not. R. Astron. Soc. 388 275
(2008).

\bibitem{w-JBP} H. K. Jassal, J. S. Bagla, and T. Padmanabhan, Mon. Not. R.
Astron. Soc. Letters 356(1) L11 (2005).

\bibitem{w-LOG} G. Efstathiou, Mon. Not. R. Astron. Soc. 310 842 (1999).

\bibitem{w-BA} E. M. Barboza, Jr. and J. S. Alcaniz, J. Cosmol. Astropart.
Phys. 02 042 (2012).

\bibitem{Liddle} A.R. Liddle, An introduction to cosmological inflation. High energy physics and cosmology, 260 (1998).

\bibitem{w-sqrt} G. Pantazis, S. Nesseris and L. Perivolaropoulos, Phys.
Rev. D 93 103503 (2016).

\bibitem{w-LIN1} D. Huterer, M. S. Turner, Phys. Rev. D 64 123527 (2001).

\bibitem{w-LIN2} J. Weller and A. Albrecht, Phys. Rev. D 65 103512 (2002).

\bibitem{w-FSSL} C. -J. Feng et al., J. Cosmol. Astropart. Phys. 09 023
(2012).

\bibitem{w-sin} R. Lazkoz, V. Salzano, and I. Sendra, Phys. Lett. B 694 198
(2010).

\bibitem{w-MZ} J. -Z. Ma and X. Zhang, Phys. Lett. B 699 233 (2011).

\bibitem{w-feng} L. Feng and T. Lu, J. Cosmol. Astropart. Phys. 11 034
(2011).

\bibitem{Mackey/2013} D. F. Mackey et al., \textit{Publ. Astron. Soc. Pac.} 
\textbf{125}, 306 (2013).

\bibitem{Yu/2018} Yu, B. Ratra, F-Yin Wang, \textit{Astrophys. J.} \textbf{856}, 3
(2018).

\bibitem{Moresco/2015} M. Moresco, \textit{Month. Not. R. Astron. Soc.} \textbf{450}, L16-L20 (2015).

\bibitem{Sharov/2018} G.S. Sharov, V.O. Vasilie, \textit{Mathematical Modelling and
Geometry} \textbf{6}, 1 (2018).

\bibitem{BAO1} C. Blake et al., \textit{ Mon. Not. Roy. Astron. Soc.} \textbf{418},
1707 (2011).

\bibitem{BAO2} W. J. Percival et al., \textit {Mon. Not. Roy. Astron. Soc.} \textbf{401%
}, 2148 (2010).

\bibitem{BAO3} F. Beutler et al., \textit{ Mon. Not. Roy. Astron. Soc.} \textbf{416},
3017 (2011).

\bibitem{BAO4} N. Jarosik et al., \textit{ Astrophys. J. Suppl.} \textbf{192}, 14
(2011).

\bibitem{BAO5} D. J. Eisenstein et al., \textit{ Astrophys. J.} \textbf{633}, 560
(2005).

\bibitem{BAO6} R. Giostri et al.,\textit{ J. Cosm. Astropart. Phys.} \textbf{1203},
027 (2012).

\bibitem{Scolnic/2018} D.M. Scolnic et al., \textit{Astrophys. J} \textbf{859}, 101
(2018).

\bibitem{Chang/2019} Z. Chang et al., Chin. Phys. C, \textbf{43}, 125102
(2019).

\bibitem{Kessler/2017} R. Kessler, D. Scolnic, Astrophys. J., \textbf{836},
56 (2017).
\bibitem{Mamon1} A. A. Mamon, \textit{Mod. Phys. Lett. A}, \textbf{33}, 1850113 (2018).

\bibitem{Mamon2} A. A. Mamon, K. Bamba and S. Das, \textit{Eur. Phys. J. C}, \textbf{77}, 1 (2017).

\bibitem{Chen1} G. Chen and B. Ratra ,\textit{PASP} 
\textbf{123}, 1127 (2011).

\bibitem{Chen2} G. Chen, S. Kumar and B. Ratra,\textit{Astrophys. J.}  \textbf{835}, 86 (2017).

\bibitem{Aubourg} E. Aubourg et al.,\textit{Phys. Rev. D} 
\textbf{92}, 123516 (2015).

\bibitem{Capozziello} S. Capozziello, R. D’Agostino and O. Luongo, \textit{Mon. Not. Roy. Astron. Soc.} 
\textbf{494}, 2576-2590 (2020).


\bibitem{Farooq} O. Farooq, et al., \textit{Astrophys. J.} 
\textbf{835}, 26–37 (2017).

\bibitem{Jesus} J.F. Jesus, et al.,\textit{J. Cosmol. Astropart. Phys.} 
\textbf{2020}, 053 (2020).

\bibitem{Garza} J.R. Garza, et al., \textit{Eur. Phys. J. C} 
\textbf{79}, 890 (2019).


\bibitem{Hernandez} A. Hernandez-Almada, et al., \textit{Eur. Phys. J. C} 
\textbf{79}, 12 (2019).

\bibitem{Zhang} Q. J. Zhang and Y. L. Wu, \textit{J. Cosmol. Astropart. Phys.} 
\textbf{2010}, 08 (2010).

\bibitem{S1} J. Zhang, X. Zhang, H. Liu, \textit{Phys. Lett. B} 
\textbf{651}, 2-3 (2007).

\bibitem{S2} L. N. Granda, \textit{Int. J. Mod. Phys. D} 
\textbf{18}, 11 (2009).

\bibitem{Sahni} V. Sahni et al., \textit{JETP Lett.} \textbf{77}, 201 (2003).

\bibitem{Alam} U. Alam et al., \textit{ Month. Not. Roy. Astron. Soc.} {\bf 344}, 1057 (2003). 

\bibitem{Sahni1} V. Sahni, A. Shafieloo, and A. A. Starobinsky, \textit{Phys. Rev. D} {\bf 78}, 103502 (2008).

\bibitem{Michel} F.C. Michel, \textit{Astrophys. Space Sci.} {\bf  15}, 153 (1972).

\bibitem{Babichev1} E. Babichev et al., \textit{Phys. Rev. Lett.} {\bf  93}, 021102 (2004).

\bibitem{Babichev2} E. Babichev, V. Dokuchaev, and Y. Eroshenko, \textit{J. Exp. Theor. Phys.} {\bf  11}, 528 (2005).

\bibitem{Sarkar} S. Sarkar, \textit{Astrophys. Space Sci.} {\bf  352}, 245-253 (2014).

\bibitem{Thorne} K. S. Thorne et al., Black holes: the membrane paradigm, \textit{Yale university press}, Black holes: the membrane paradigm (1986).

\bibitem{Haw1} S. Hawking, Black holes and the information paradox? \textit{In General Relativity and Gravitation} (2005).

\bibitem{Bek} J. D. Bekenstein, \textit{Phys. Today} {\bf 33}, 24-31 (1980).

\bibitem{Haw2} S. Hawking, \textit{Nature} {\bf  248}, 30-31 (1974).

\bibitem{Cline} D.B. Cline, D.A. Sanders,W. Hong, \textit{Astrophys. J.} {\bf  486}, 169 (1997).

\end{thebibliography}
\end{document}